\documentclass[journal,draftclsnofoot,onecolumn,12pt,twoside]{IEEEtran}

\usepackage[T1]{fontenc}

\usepackage{cite}
\usepackage{amsmath,amssymb,amsfonts}
\usepackage{algorithm}
\usepackage{algorithmic}
\usepackage{graphicx}
\usepackage{textcomp}
\usepackage{xcolor}
\usepackage{subfigure}
\usepackage{booktabs}
\usepackage{multirow}
\usepackage{url}
\makeatletter
\def\UrlAlphabet{%
      \do\a\do\b\do\c\do\d\do\e\do\f\do\g\do\h\do\i\do\j%
      \do\k\do\l\do\m\do\n\do\o\do\p\do\q\do\r\do\s\do\t%
      \do\u\do\v\do\w\do\x\do\y\do\z\do\A\do\B\do\C\do\D%
      \do\E\do\F\do\G\do\H\do\I\do\J\do\K\do\L\do\M\do\N%
      \do\O\do\P\do\Q\do\R\do\S\do\T\do\U\do\V\do\W\do\X%
      \do\Y\do\Z}
\def\UrlDigits{\do\1\do\2\do\3\do\4\do\5\do\6\do\7\do\8\do\9\do\0}
\g@addto@macro{\UrlBreaks}{\UrlOrds}
\g@addto@macro{\UrlBreaks}{\UrlAlphabet}
\g@addto@macro{\UrlBreaks}{\UrlDigits}
\makeatother

\graphicspath{{figures/}}
\def\tr{\mathrm{tr}}
\def\re{\mathrm{Re}}
\def\im{\mathrm{Im}}
\def\OAMP{\text{OAMP}}
\def\MAMP{\text{MAMP}}
\def\CGOAMP{\text{CG-OAMP}}
\def\BO{{Bayes-optimal}}
\def\Tx{{Tx}}
\def\Rx{{Rx}}
\newcommand{\Times}[2]{${\text{#1}\times\text{#2}}$}
\newcommand{\SNR}[2]{${\text{SNR} #1 \text{#2}\;\text{dB}}$}

\begin{document}
\ifdefined \GramaCheck
  \newcommand{\CheckRmv}[1]{}
  \newcommand{\figref}[1]{Figure 1}%
  \newcommand{\tabref}[1]{Table 1}%
  \newcommand{\secref}[1]{Section 1}
  \newcommand{\algref}[1]{Algorithm 1}
  \renewcommand{\eqref}[1]{Equation 1}
\else
  \newcommand{\CheckRmv}[1]{#1}
  \newcommand{\figref}[1]{Fig.~\ref{#1}}%
  \newcommand{\tabref}[1]{Table~\ref{#1}}%
  \newcommand{\secref}[1]{Sec.~\ref{#1}}
  \newcommand{\algref}[1]{Algorithm~\ref{#1}}
  \renewcommand{\eqref}[1]{(\ref{#1})}
\fi

%
\title{Model-Driven Deep Learning-Based MIMO-OFDM Detector: 
Design, Simulation, and Experimental Results}
%
%
%

\author{Xingyu~Zhou,
        Jing~Zhang,
		Chen-Wei~Syu,	
		Chao-Kai~Wen,
		Jun~Zhang,
        and~Shi~Jin
\thanks{Xingyu Zhou, Jing Zhang, and Shi Jin are with the National Mobile
Communications Research Laboratory, Southeast University, Nanjing 210096, China
(e-mail: xy\underline{  }zhou@seu.edu.cn; jingzhang@seu.edu.cn; jinshi@seu.edu.cn).}
\thanks{Chen-Wei Syu and Chao-Kai Wen are with Institute of Communications Engineering,
National Sun Yat-sen University, Kaohsiung 80424, Taiwan
(e-mail: syuwei110014@gmail.com; chaokai.wen@mail.nsysu.edu.tw).}
\thanks{Jun Zhang is with College of Telecommunications and Information Engineering,
Nanjing University of Posts and Telecommunications, Nanjing 210003, China
(e-mail: zhangjun@njupt.edu.cn).}
\thanks{This paper was presented in part at the IEEE International
Conference on Communications (ICC) Workshop, Montreal,
Canada, Jun. 2021 \cite{zhou_model-driven_2021}.}}


%


\maketitle
\vspace{-1.8cm}

\begin{abstract}
\vspace{-0.2cm}
Multiple-input multiple-output orthogonal frequency division multiplexing (MIMO-OFDM),
a fundamental transmission scheme, promises high throughput and robustness against
multipath fading. 
However, these benefits rely on the efficient detection strategy at the receiver and
come at the expense of the extra bandwidth consumed by the cyclic prefix (CP).
We use the iterative orthogonal approximate message passing (OAMP) algorithm in this paper
as the prototype of the detector because of its remarkable potential for interference suppression.
However, OAMP is computationally expensive for the matrix inversion per iteration. 
We replace the matrix inversion with the conjugate gradient (CG) method to reduce the
complexity of OAMP. 
We further unfold the CG-based OAMP algorithm into a network and tune the critical parameters 
through deep learning (DL) to enhance detection performance.
Simulation results and complexity analysis show that the proposed scheme has significant
gain over other iterative detection methods and exhibits comparable performance to 
the state-of-the-art DL-based detector at a reduced computational cost. 
Furthermore, we design a highly efficient CP-free MIMO-OFDM receiver architecture 
to remove the CP overhead.
This architecture first eliminates the intersymbol interference by buffering the
previously recovered data and then detects the signal using the proposed detector. 
Numerical experiments demonstrate that the designed receiver offers a higher spectral
efficiency than traditional receivers. 
Finally, over-the-air tests verify the effectiveness and robustness of the
proposed scheme in realistic environments.

\end{abstract}

\vspace{-0.6cm}

%
\IEEEpeerreviewmaketitle


\section{Introduction}
%
%
%
%
Multiple-input multiple-output orthogonal frequency division multiplexing (MIMO-OFDM), 
as a combination of multi-antenna and multi-carrier technologies, has shown its remarkable 
potential for high-speed transmission in current and future wireless communication systems
\cite{van_zelst_implementation_2004}.
However, the system design for MIMO-OFDM faces many challenges. 
{Efficient detectors are essential as the number of antennas increases in modern communications.}
Another major challenge is the redundancy issue. 
In the standard OFDM scheme, a cyclic prefix (CP) is inserted between OFDM blocks to
mitigate the intersymbol interference (ISI),  
which results in degenerated spectral efficiency because this CP comprises
redundant data \cite{goldsmith_wireless_2005}. 

A tradeoff between complexity and performance should be achieved in the receiver design
for MIMO-OFDM systems considering detection strategy \cite{wu_low-complexity_2014}.
The optimal maximum likelihood (ML) detection has a prohibitive complexity exponentially growing
with the number of decision variables 
and thus cannot be applied in practice.
The suboptimal zero-forcing and linear minimum mean square error (LMMSE) detectors 
reduce the complexity to an acceptable level but suffer a large performance gap
compared to the optimality.
The iterative approximate message passing (AMP) \cite{donoho_message-passing_2009} algorithm  
has been recently applied in MIMO detection due to its low complexity \cite{wu_low-complexity_2014}. 
AMP uses the central limit theorem and Taylor series expansion \cite{montanari2011graphical} 
to approximate traditional message passing and significantly reduces the complexity. 
AMP is proven to be Bayes-optimal for large (i.e., the system dimensions grow to infinity with a fixed ratio) 
independent and identically distributed (IID) sub-Gaussian channel matrices.   
However, AMP becomes unstable when the channel matrix deviates from the strong 
assumption of IID sub-Gaussian.
Orthogonal AMP (OAMP) has been proposed to relax the constraint on the matrix to the
general unitarily-invariant matrices \cite{ma_orthogonal_2017}. 
However, OAMP involves a direct matrix inversion per iteration, rendering
the algorithm computationally expensive. 
Several techniques, such as the convolutional AMP (CAMP) \cite{takeuchi_bayes-optimal_2021-1} 
and the memory AMP (MAMP) \cite{liu_memory_2021}, have emerged to solve this limitation.
The two algorithms have a similar idea in replacing the high-complexity LMMSE estimator 
in OAMP with a low-complexity long-memory matched filter. 
CAMP and MAMP inherit the low-cost strength of AMP while preserving the {\BO}
property for unitarily-invariant matrices.
However, all the above-mentioned AMP-type detectors generally have weaknesses, which lie in
the difficult satisfaction of their prerequisites in practice. 
The large system limit and the assumption on channel matrices no longer hold when faced
with realistic small- or medium-sized (e.g., 8 $\times$ 8 and 32 $\times$ 32) MIMO systems 
or channels with strong correlations, and the performance of the detectors severely deteriorates.

Moreover, the design of a CP-free MIMO-OFDM system is attractive for resolving the redundancy issue.
However, the absence of CP breaks the orthogonality between adjacent subcarriers and leads to 
severe ISI and intercarrier interference (ICI), thus complicating data detection.  
A series of studies have investigated the techniques to tackle this problem 
\cite{aminjavaheri_ofdm_2017,dukhyun_kim_residual_1998,liu_symbol_2017,liu_successive_2019,pham_channel_2017}.
The mathematical analysis of CP removal in massive MIMO-OFDM systems is investigated 
in \cite{aminjavaheri_ofdm_2017}, which reveals that the ISI and ICI do not fade away 
despite the infinite growth of the number of base station (BS) antennas.
An iterative strategy combining tail cancellation and cyclic restoration is proposed 
in \cite{dukhyun_kim_residual_1998} to cancel the residual ISI.
However, the algorithm is fragile considering the channel impulse response length
and seriously deteriorates when faced with long delay spreads. 
The authors of \cite{liu_symbol_2017,liu_successive_2019} modified this technology by 
using decision and stored feedback equalizations to remove ISI.
Moreover, a trellis equalizer is developed in \cite{pham_channel_2017} to combat the interference
generated by insufficient CP in MIMO-OFDM systems. However, the technique entails
a multistep detection process with high complexity.

Owing to its overwhelming privilege in finding data representation, deep learning (DL)
has been widely used in physical layer communication recently \cite{qin_deep_2019} 
and has emerged to provide a different approach for solving traditional challenging problems 
\cite{ye_power_2018,gao_comnet_2018,zhang_artificial_2019}. 
A fully connected deep neural network (FC-DNN), which is more robust than traditional
methods in CP-free or pilotless situations, is proposed to process channel 
estimation and data detection for an OFDM system \cite{ye_power_2018}. 
However, this approach treats the receiver as a black box and relies on extensive data and training time.
As an alternative technology, model-driven DL \cite{xu_model-driven_2018} integrates 
domain knowledge into neural network (NN) design and maintains the block-based structure
of the communication system, which significantly reduces the training cost. 
In particular, the authors of \cite{gao_comnet_2018} proposed a model-driven OFDM receiver named ComNet.
The ComNet receiver, which combines a recurrent NN with communication intelligence,  
is superior to the FC-DNN in deployment efficiency and performance. 
However, these works, which are developed for single-input single-output (SISO)-OFDM, 
can be complicated when directly extended to MIMO-OFDM. 

DL-based MIMO detectors have also achieved promising results on balancing detection accuracy and complexity 
\cite{samuel_learning_2019,khani_adaptive_2020,ito_trainable_2019,he_model-driven_2020}.
A deep MIMO detection network called DetNet is developed in \cite{samuel_learning_2019} utilizing the idea of algorithm unfolding \cite{monga_algorithm_2021}. 
The DetNet has near-optimal detection performance and comparable complexity as AMP under
well-conditioned channels but cannot deal with correlated channels.
An NN named MMNet is constructed in \cite{khani_adaptive_2020} based on the iterative soft-thresholding algorithm (ISTA). 
MMNet can handle remarkably challenging scenarios by adding a trainable matrix of parameters in the network.  
However, DetNet and MMNet have excessive parameters to be optimized, 
contributing to the reduced efficiency of the training process.
DNN-aided message passing and soft interference cancellation detectors are respectively
introduced in \cite{tan_improving_2020} and \cite{shlezinger2021deepsic} to improve robustness
against varying antenna configurations and channel model uncertainty.
A trainable ISTA (TISTA) network, which can also be viewed as model-driven, 
is proposed in \cite{ito_trainable_2019} for signal reconstruction problems. 
The TISTA unfolds the iterations of OAMP into several layers and requires only one adjustable
parameter for each layer, which leads to a stable and fast training process. 
Inspired by the TISTA, the authors of \cite{he_model-driven_2020} introduced a few
additional parameters into the OAMP iteration and derived another NN called OAMP-NET
for data detection in MIMO systems and CP-free OFDM channels \cite{zhang_artificial_2019}. 
OAMP-NET possesses more flexibility than TISTA and can be rapidly trained and deployed.
However, the OAMP-NET has not resolved the inherent high-complexity problem of OAMP.

We develop a model-driven NN for MIMO-OFDM signal detection in this paper based on the
aforementioned studies. 
The prototype of the network is the iterative OAMP detector, which has a strong capability
to suppress interference but involves high complexity.
We initially introduce the conjugate gradient (CG) method 
to avoid direct matrix inversion in OAMP and derive a CG-based OAMP (CG-OAMP) detector 
\cite{takeuchi_rigorous_2017}.
We further unfold the CG-OAMP detector into a NN, namely CG-OAMP-NET, which is inspired by 
the TISTA in \cite{ito_trainable_2019} and OAMP-NET in \cite{he_model-driven_2020}, because
conventional AMP-type algorithms suffer performance degradation in realistic finite-dimensional MIMO systems.
The detection performance can be significantly improved by tuning a few key parameters in the network through DL.  
Moreover, a CG-OAMP-NET-based CP-free MIMO-OFDM receiver is proposed to handle 
the difficulty engendered by the lack of CP and improve spectral efficiency.
The contributions of this paper can be summarized as follows:
\begin{itemize}
\item We propose a model-driven CG-OAMP-NET detector, which combines the benefits of domain knowledge and DL. 
In comparison with the OAMP-NET in \cite{he_model-driven_2020}, the proposed CG-OAMP-NET uses
CG to alleviate the high-complexity disadvantage of OAMP without compromising the performance.
Complexity analysis demonstrates that the running time of CG-OAMP-NET is substantially lower than 
those of OAMP and OAMP-NET.
Meanwhile, we tune the trainable parameters in the network {to improve the detection performance} using the DL.
Simulation results indicate that CG-OAMP-NET has considerable gain over the prototype
OAMP/CG-OAMP and low-cost iterative algorithms, such as MAMP. 
Thus, CG-OAMP-NET can achieve a desirable tradeoff between detection performance and complexity.
\item 
We apply the proposed CG-OAMP-NET to the design of a CP-free MIMO-OFDM receiver. In combination with the 
residual ISI elimination strategy, we derive a highly efficient receiver architecture,
which can effectively combat the interference caused by CP removal and offer a higher spectral efficiency than 
traditional receivers with sufficient CP.
\item Exceeding beyond our early studies in \cite{zhou_model-driven_2021}, 
we validate the proposed scheme with a wide range of numerical experiments and further build
an over-the-air (OTA) platform to demonstrate its performance 
in realistic environments because real-life channels can be more complicated than the models used in simulations. 
Experimental results show that the scheme is robust to practical environments
and is feasible and promising for future communications.
\end{itemize}

\textit{Notations:}
Boldface letters denote column vectors or matrices. 
$\mathbf{A}^T$, $\mathbf{A}^{\mathrm{H}}$, and $\tr(\mathbf{A})$ represent the transpose, conjugate transpose, 
and trace of matrix $\mathbf{A}$, respectively. 
$\re ( \cdot )$ and  $\im ( \cdot )$ respectively represent the real and imaginary parts of a complex value. 
Furthermore, ${\mathbf{I}}_N$ is an ${N\times N}$ identity matrix, and
$\mathbf{0}$ is a zero matrix. 
$\otimes$, $\left \| \cdot \right \|$, and ${\bf{E}}\{ \cdot \} $ denote the Kronecker
product, Euclidean norm, and expectation operation, respectively. 
Finally, ${\cal N}\left( {z;\mu ,{\sigma ^2}} \right)$ and ${\cal CN}\left( {z;\mu ,{\sigma ^2}} \right)$ 
denote real- and complex-valued Gaussian random variables with mean $\mu $ and variance $\sigma^2$, respectively.


\section{SYSTEM MODEL} \label{sec:sys_model}

\CheckRmv{
\begin{figure}[tbp]
	\centering
	\subfigure[Transmitter]{
		\label{fig:TX_and_channel}
		\includegraphics[width=4.3in]{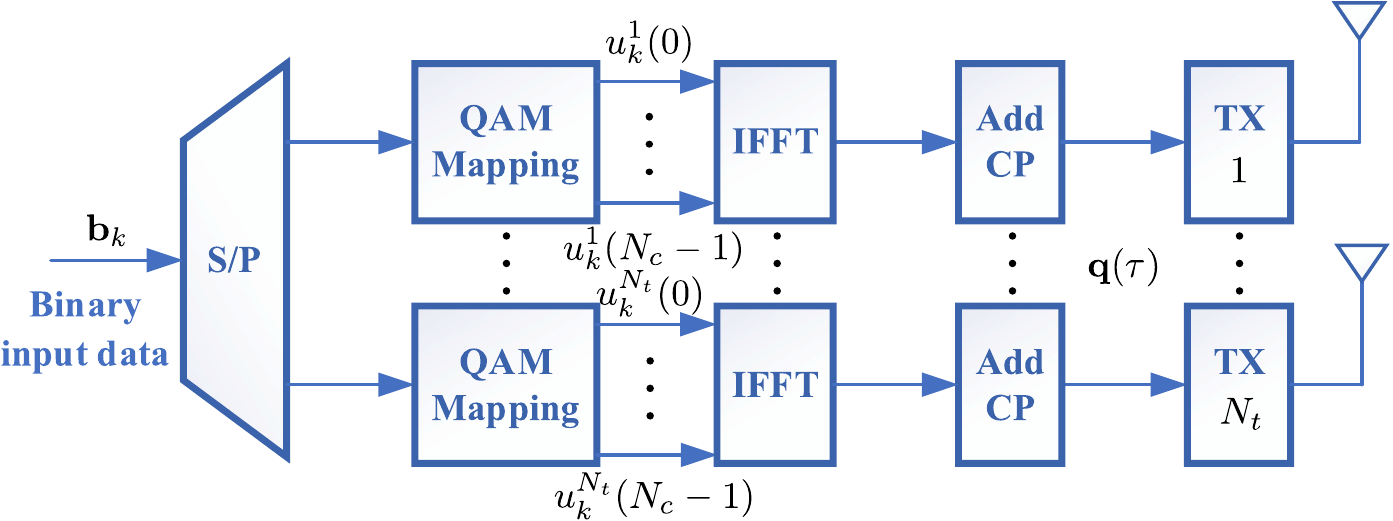}
	}
	\subfigure[Receiver]{
		\label{fig:Rx}
		\includegraphics[width=4.5in]{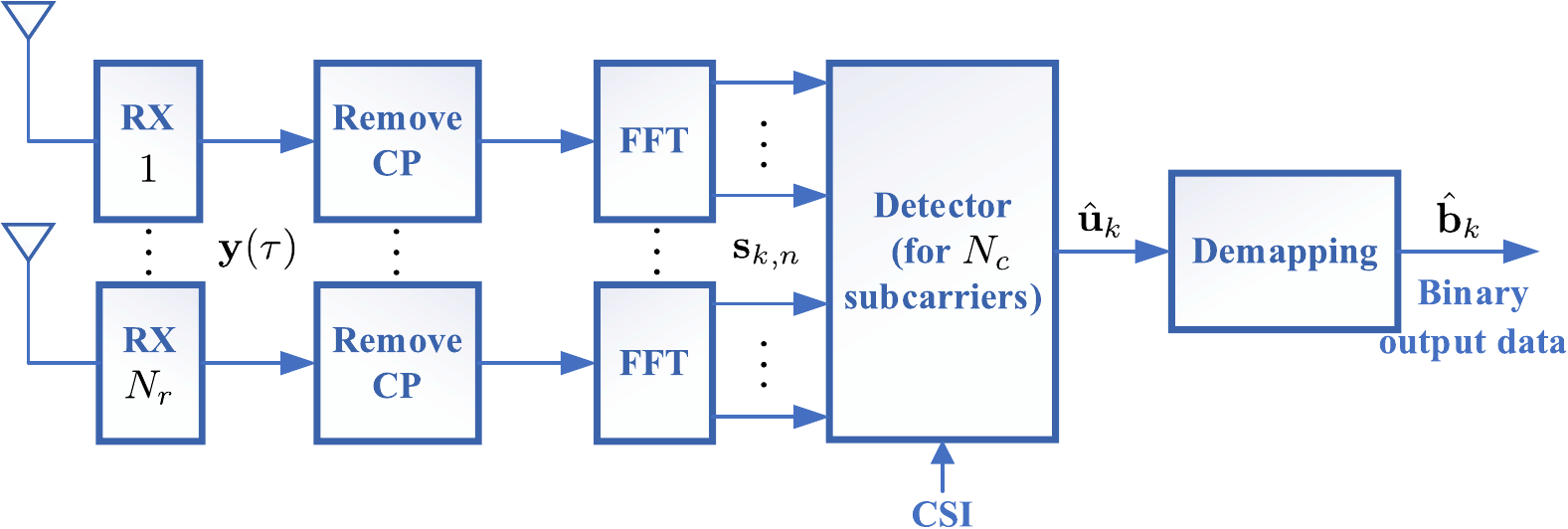}
	}
	\caption{Block diagram of an uncoded MIMO-OFDM system.}
	\label{fig:MIMO_OFDM_system}
\end{figure}
}

We consider {the downlink\footnote{{Conventional linear and AMP-type detectors fail when faced with a small Rx-to-Tx-antenna ratio $N_r/N_t$, which is a common case in the downlink. The main motivation of this work is to use DL to overcome the performance degradation in this situation. Thus, we consider the downlink transmissions and assume that $N_r/N_t$ is close to one to avoid the precoding design.
}} of} an uncoded MIMO-OFDM system with $N_t$ transmit (\Tx) and $N_r$ receive (\Rx) antennas. 
As shown in \figref{fig:MIMO_OFDM_system}, the system contains a transmitter and a receiver. 
The transmitter includes $N_t$ OFDM {\Tx} branches, and each branch utilizes $N_c$ subcarriers for transmission.\footnote{{The power allocation process is omitted, and we assume a uniform power allocation over the subcarriers. In practical systems, the power allocation can be solved by the water-filling algorithm to approach the capacity of multipath channels \cite{brandenburg_capacity_1974}.}} 
The input bits are denoted as $\mathbf{b}_k$ for the $k$-th ($k = {0},{1}, \dots$) MIMO-OFDM symbol duration.
The incoming bits are first multiplexed to each branch and then converted to quadrature 
amplitude modulation (QAM) symbols independently. 
The frequency-domain symbol sequence transmitted by the $p$-th branch for the $k$-th duration
is denoted as $\{ u_k^p(n)\}_{n=0}^{N_c-1}$, where each symbol is drawn from the $P$-QAM 
constellation ${\mathcal{A} = \{ {a_1},\dots,{a_P}\}}$. 
$\{ u_k^p(n)\}_{n=0}^{N_c-1}$ is then transformed into the time domain by an $N_c$-point 
inverse fast Fourier transform (IFFT), and a CP with $N_g$ time samples is inserted 
before transmission in the channel.
We consider a quasi-static wireless channel that remains unchanged during a symbol duration. 
A finite impulse response filter with $L$ filter taps, namely $h_{k,l}^{qp},l \in \{ {0}, \dots, L-{1}\}$,
is utilized to model the discrete multipath channel between the $p$-th {\Tx} and $q$-th {\Rx} antenna. 
The CP length should be longer than the channel delay
spread, namely $N_g\geq L-{1}$.

The block diagram of the receiver is presented in \figref{fig:Rx}. 
Let ${\mathbf{q}(\tau) \in \mathbb{C}^{N_t \times 1}}$ denote the MIMO vector sent
at time instance $\tau$.
The corresponding received $N_r$-dimensional MIMO vector $\mathbf{y}(\tau)$ is then given by 
\CheckRmv{
\begin{equation}
	\mathbf{y}(\tau) = \sum_{l=0}^{L-1}\mathbf{H}_{k,l} \mathbf{q}(\tau-l)
	+ \mathbf{w}(\tau), 
\end{equation}
}
where $\mathbf{w}(\tau)$ is the additive white Gaussian noise (AWGN) 
whose elements are independent of each other and follow $\mathcal{CN}(0,\sigma_w^2)$ 
with $\sigma_w^2$ as the noise variance.
$\mathbf{H}_{k,l} \in \mathbb{C}^{N_r \times N_t}$ is the  MIMO channel matrix 
on the $l$-th path for the $k$-th MIMO-OFDM symbol, and the $(q,p)$-th element of $\mathbf{H}_{k,l}$ is given by $h_{k,l}^{qp}$.
The resulting signal after CP removal is sent to the FFT modules to recover in the frequency domain
and yields a flat-fading signal model as follows: 
\CheckRmv{
\begin{equation}
	\mathbf{s}_{k,n} = \mathbf{G}_{k,n} \mathbf{u}_{k,n}+ \boldsymbol{\nu}_{k,n},
	\label{eq:MIMO_model}
\end{equation}
}
where ${\mathbf{u}_{k,n}=[u_k^1(n), \dots ,u_k^{N_t}(n)]^T \in \mathbb{C}^{N_t\times 1}}$, 
${\mathbf{s}_{k,n}\in \mathbb{C}^{N_r\times 1}}$, and $\boldsymbol{\nu}_{k,n}$ denote 
the {\Tx} signal, received signal, and noise vectors on the $n$-th subcarrier of the $k$-th duration, respectively.
${\mathbf{G}_{k,n}}$ is the frequency-domain channel matrix and given by 
${\mathbf{G}_{k,n}= \sum_{l ={0}}^{L -{1}} {\mathbf{H}_{k,l}\exp (-j2\pi \frac{nl}{N_c})}}$.
We assume that perfect channel state information (CSI) is known at the receiver { without specific instructions} 
because we focus on signal detection to recover ${\mathbf{u}}_{k,n}$ in this paper.  
The detector generates the estimation $\hat{\mathbf{u}}_k$ based on the CSI,
which is investigated in Section \ref{sec:CG_OAMP_NET}.
Finally, $\hat{\mathbf{u}}_k$ is fed to the demapping module, and the $N_t$ recovered data
streams are combined as the binary output $\hat{\mathbf{b}}_k$.

\section{CG-OAMP-NET} \label{sec:CG_OAMP_NET}
We develop a model-driven DL-based detector for MIMO-OFDM in this section.
First, we review the OAMP algorithm and introduce its variant, namely the low-complexity CG-OAMP
\cite{takeuchi_rigorous_2017}.
Then, we present the proposed CG-OAMP-NET in detail and further apply the detector to a CP-free scenario.
Finally, we provide the complexity analysis of the proposed detector.

\subsection{Low-Complexity CG-OAMP Detector}

The DL technique is introduced to improve detection performance, 
which is always conducted in the real-valued domain. Thus, we first perform a real-valued
decomposition on \eqref{eq:MIMO_model}\footnote{For brevity, the indexes $k$ and $n$ are dropped.} 
\CheckRmv{
\begin{equation}
	\bar{\mathbf{s}} = \bar{\mathbf{G}}\bar{\mathbf{u}} + \bar{\boldsymbol{\nu}},
\label{eq:real_domain_MIMO_model}
\end{equation}
}
where \[
	\bar{\mathbf{s}} = {\big[ {{\re} {{({\mathbf{s}})}^T},{\im} 
	{{({\mathbf{s}})}^T}}\big]^T}, 
	\bar{\mathbf{u}} = {\big[ {{\re} {{({\mathbf{u}})}^T},{\im} 
	{{({\mathbf{u}})}^T}} \big]^T},
\]
\[
	\bar{\boldsymbol{\nu}} = {\big[ {{\re} {{(\boldsymbol{\nu})}^T},{\im}
	 {{(\boldsymbol{\nu})}^T}} \big]^T},
	\bar{\mathbf{G}} = \left[ 
	\begin{array}{ccc}
		{{\re} ({\mathbf{G}})} & { - {\im} ({\mathbf{G}})}\\
		{{\im} ({\mathbf{G}})} & {{\re} ({\mathbf{G}})}
	\end{array} 
	\right].
\]

We can use the OAMP algorithm \cite{ma_orthogonal_2017} to detect the transmitted symbol $\bar{\mathbf{u}}$ in \eqref{eq:real_domain_MIMO_model} based on the known 
$\bar{\mathbf{s}}$, $\bar{\mathbf{G}}$, and noise variance $\sigma_{\nu}^2$. 
OAMP comprises a linear estimator (LE) and a nonlinear estimator (NLE).
The algorithm iteratively exchanges information between the two local estimators until convergence. 
Initialized with $\hat{\mathbf{u}}_1 = \mathbf{0}$, the iterative process proceeds as follows:
\CheckRmv{
	\begin{align}
	&\text{LE:}\quad ~~\, {\mathbf{r}}_t = \hat{ \mathbf{u}}_t +{\mathbf{W}}_t
	\left ({\bar{\mathbf {s}}-\bar{\mathbf{G}}\hat{\mathbf{u}}_t}\right ),\label{eq:LE}\\
	&\text{NLE:}\quad  \hat{\mathbf{u}}_{t + 1} = \eta _{t} \left (
	{ \mathbf{r}}_t \right ), \label{eq:NLE}
\end{align}
}
where the subscript $t$ is the iteration time index, and the final estimation is 
$\hat{\mathbf{u}}_{T + 1}$ with $T$ as the total number of iterations.
Two error vectors, namely $\mathbf{e}_t=\mathbf{r}_t-\bar{\mathbf{u}}$ and 
$\mathbf{f}_t=\hat{ \mathbf{u}}_t-\bar{\mathbf{u}}$, are introduced to evaluate the accuracy
of the estimators. The error variance estimators are defined as
\CheckRmv{
	\begin{equation}
		\tau _t^2 = \frac{{1}}{{2}N_t}{\bf{E}}\{ {\left\| {{{\bf{e}}_t}} \right\|^2}\},
		v_{t+1}^2 = \frac{{1}}{{2}N_t}{\bf{E}}\{ {\left\| {{{\bf{f}}_t}} \right\|^2}\}. 
	\end{equation}	
}

The LE is specified by the de-correlated matrix $\mathbf{W}_t$, which satisfies
${\tr(\mathbf{I}-\mathbf{W}_t\bar{\mathbf{G}})={0}}$. For any given matrix $\hat{\mathbf{W}}_t$
with ${2}N_t \times {2}N_r$ dimension, $\mathbf{W}_t$ can be constructed as follows:
\CheckRmv{
\begin{equation}
	{\mathbf{W}}_t = \zeta_t \hat{\mathbf{W}}_t, \label{eq:W}
\end{equation}
}
where the de-correlated coefficient $\zeta_t = \frac{{2}N_t}{\tr(\hat{\mathbf{W}}_t\bar{\mathbf{G}})}$ 
is vital for maintaining the orthogonality between estimation errors.
Furthermore, the {\BO} $\hat{\mathbf{W}}_t$ takes an LMMSE structure.
\CheckRmv{
\begin{equation}
	\hat{\bf{W}}_t = \bar{\bf{G}}^T \left( \bar{\bf{G}} \bar{\bf{G}}^T 
	+ \frac{\sigma _{\nu}^2}{{2}v_{t}^2}{\bf{I}} \right)^{ - 1}. \label{eq:LMMSE}
\end{equation}
}

The NLE $\eta_t(\cdot)$ in \eqref{eq:NLE} is required as an element-wise divergence-free 
function, which is discussed in \cite{ma_orthogonal_2017}. However, some 
parameters in constructing the {\BO} divergence-free function are related to the prior
signal distribution and difficult to compute \cite{he_model-driven_2020}. 
Therefore, we use an easy-to-implement MMSE estimator.
\CheckRmv{
\begin{equation}
	\hat{\mathbf{u}}_{t + 1} = {\bf{E}}\{\bar{\mathbf{u}}| \mathbf{r}_t,\tau_t\}.\label{eq:MMSE} 
\end{equation}
} 
The MMSE estimation for each element ${\bar{u}}^{n}$ of $\bar{\mathbf{u}}$ can be calculated as 
\CheckRmv{
\begin{equation}
	{\bf{E}}\{{\bar{u}}^{n}| r_{t}^{n},\tau_t\}=\sum_{a_{m} \in \mathcal{\bar A}} a_{m} 
	\frac{\mathcal{N}(a_{m} ; r_{t}^{n}, \tau_{t}^{2})}{\sum_{a_{m} \in \mathcal{\bar A}} 
	\mathcal{N}(a_{m} ; r_{t}^{n}, \tau_{t}^{2})},
\end{equation}
}
where ${\mathcal{\bar A}=\{a_1,\dots,a_{\sqrt{P}}\}}$ is the real component of the $P$-QAM modulation set.

According to \cite[(30),(31)]{ma_orthogonal_2017}, $\tau_t^2$ and $v_{t+1}^2$ can be estimated as follows:
\CheckRmv{
\begin{align}
	\tau_t^2 &= \frac{{1}}{{2}N_t}\tr(\mathbf{B}_t\mathbf{B}_t^T)v_t^2+
	\frac{1}{{4}N_t}\tr(\mathbf{W}_t\mathbf{W}_t^T)\sigma_{\nu}^2,\label{eq:tau2_old}\\
	v_{t+1}^2 &= \frac{ {\| \bar{\bf{s}} - \bar{\bf{G}} \hat{\bf{u}}_{t+1}\|}^2 - N_r\sigma_{\nu}^2}
	{{\tr}( \bar{\bf{G}}^T \bar{\bf{G}} )}, \label{eq:v2}
\end{align}
}
where $\mathbf{B}_t=\mathbf{I}-\mathbf{W}_t\bar{\mathbf{G}}$. Given the definition of $\mathbf{W}_t$
in \eqref{eq:W} and \eqref{eq:LMMSE},
\eqref{eq:tau2_old} can be written as follows for easy calculation. 
\CheckRmv{
\begin{equation}
	\tau _t^2 = v_t^2 ( \zeta_t - {1} ). \label{eq:tau2}
\end{equation}
} 
We provide the derivation of \eqref{eq:tau2} in the Appendix.
For $v_{t+1}^2$ in \eqref{eq:v2}, we smooth the update through a convex combination of the former value
to strengthen the algorithm \cite{cespedes_expectation_2014}, that is,
\CheckRmv{
\begin{equation}
	v_{t+1}^2 \leftarrow \beta v_{t+1}^2 + ( {1}-\beta )v_t^2, \label{eq:damping}
\end{equation}
}   
where $\beta \in [0,1]$ is the damping factor and is selected as 0.5 in the proposed implementation.
Furthermore,  $v_{t+1}^2$ is replaced by $\max(v_{t+1}^2,\epsilon)$, 
where $\epsilon$ is a small positive constant and is set as $\epsilon={1}\times{10}^{-10}$
in the experiments, to avoid stability problems.

Calculating the LMMSE estimation \eqref{eq:LMMSE} causes the complexity of $\mathcal{O}(N_r^3)$ 
because it needs a matrix inversion with $2N_r \times 2N_r$ dimension, which is time-consuming. 
Hence, a low-complexity CG-based implementation of \eqref{eq:LMMSE} is introduced.

Substituting \eqref{eq:W} and \eqref{eq:LMMSE} into \eqref{eq:LE}, we can rewrite the LE as
\CheckRmv{
\begin{align}
	{{\mathbf{r}}_t} &= {\hat{\mathbf{u}}_t} + \zeta_t{\bar{\mathbf{G}}^T}{\left( {\bar{\mathbf{G}}
	{{\bar{\mathbf{G}}}^T} + \frac{{\sigma_{\nu}^2}}{{2v_{t}^2}}{\mathbf{I}}} \right)^{ - 1}}
	\bigg( {\bar{\mathbf{s}} - \bar{\mathbf{G}}{{\hat{\mathbf{u}}}_t}} \bigg) \nonumber \\
	&= {\hat{\mathbf{u}}_t} + \zeta_t{\bar{\mathbf{G}}^T}{\boldsymbol{\Xi}}_t^{ - 1}{\mathbf{g}}_t, \label{eq:LE_full}
\end{align}
}
where 
\CheckRmv{
\begin{align}
	{\boldsymbol{\Xi }}_t &= \bar{\mathbf{G}} \bar{\mathbf{G}}^T + \frac{\sigma_{\nu}^2}{2v_{t}^2}{\mathbf{I}}, \label{eq:Xi}\\
	{\mathbf{g}}_t &= \bar{\mathbf{s}} - \bar{\mathbf{G}} \hat{\mathbf{u}}_t. \label{eq:yt}
\end{align}
}
Let ${\mathbf{z}}_t = {\boldsymbol{\Xi}}_t^{ - 1}{\mathbf{g}}_t$, and then \eqref{eq:LE_full} is converted into
\CheckRmv{
\begin{equation}
	{{\mathbf{r}}_t} = {\hat{\mathbf{u}}_t} + {\zeta _t}{\bar{\mathbf{G}}^T}{{\mathbf{z}}_t}. \label{eq:LE_revised}
\end{equation}
}
${\mathbf{z}}_t \in \mathbb{R}^{2N_r\times 1}$ is the solution of
the symmetric positive definite linear system ${\boldsymbol{\Xi}}_t{\mathbf{z}}_t = {\mathbf{g}}_t$  
and can be solved by CG without matrix inversion \cite{takeuchi_rigorous_2017}.
\algref{alg:CG} shows that 
CG is an iterative method for approximating the exact solution of the linear system,
where ${\mathbf{x}}_i$ is the approximate solution in the $i$-th iteration.
Meanwhile, ${{\boldsymbol{\rho }}_i={\mathbf{g}}_t-{\boldsymbol{\Xi }}_t {\mathbf{x}}_i}$ and ${\mathbf{p}}_i$ 
denote the $i$-th residual vector and conjugate direction, respectively. 
{The algorithm progresses until the maximum number of iterations $I_{\textrm{CG}}$ is reached, or the norm of the residual is less than $\delta$. In our work, we set the error tolerance as $\delta = 10^{-4}$ following the choice in \cite{takeuchi_rigorous_2017} to avoid redundant iterations that have little influence on the accuracy of the solution.}

\renewcommand{\algorithmicrequire}{\textbf{Initialize:}}
\renewcommand{\algorithmicensure}{\textbf{Output:}}
\newcommand{\IfThen}[2]{
  \STATE \algorithmicif\ #1\ \algorithmicthen\ #2}
\CheckRmv{
\begin{algorithm}[tbp]
	\caption{Conjugate Gradient algorithm}
	\label{alg:CG}
	\begin{algorithmic} 
		\REQUIRE ${{\mathbf{x}}_0} = {\mathbf{0}},\;{{\boldsymbol{\rho}}_0} 
		= {\mathbf{g}}_t,\; {{\mathbf{p}}_0} = {{\boldsymbol{\rho}}_0}.$
		\FOR{$i=1$ to $I_{\textrm{CG}}$} 
			\STATE 1: Update the approximate solution ${\mathbf{x}}_{i}$,
			\begin{align*}
				\alpha_{i - 1} &= {\boldsymbol{\rho}}_{i - 1}^T {\boldsymbol{\rho}}_{i - 1}/{\mathbf{p}}_{i - 1}^T
				{\boldsymbol{\Xi}}_t{{\mathbf{p}}_{i - 1}},\\
				{{\mathbf{x}}_i} &= {{\mathbf{x}}_{i - 1}} + {\alpha_{i - 1}}{{\bf{p}}_{i - 1}}.
			\end{align*} \\
			 2: Update the residual ${\boldsymbol{\rho }}_i$ and the conjugate direction ${\mathbf{p}}_i$,
			\begin{align*}
				{\boldsymbol{\rho}}_i &= {{\boldsymbol{\rho}}_{i - 1}} - {\alpha_{i - 1}}{\boldsymbol{\Xi }}_t
				{{\mathbf{p}}_{i - 1}},\\
				\beta_{i - 1} &= {\boldsymbol{\rho}}_i^T{{\boldsymbol{\rho}}_i}/{\boldsymbol{\rho}}_{i - 1}^T
				{{\boldsymbol{\rho}}_{i - 1}},\\
				{{\mathbf{p}}_i} &= {{\boldsymbol{\rho}}_i} + {\beta_{i - 1}}{{\mathbf{p}}_{i - 1}}.
			\end{align*} \\
			\IfThen {$\| {\boldsymbol{\rho}}_i \|<{\delta}$}
      			{\textbf{quit}}
		\ENDFOR
		\ENSURE ${\mathbf{z}}_t={\mathbf{x}}_{i}$
	\end{algorithmic} 
\end{algorithm}
}

The CG method has complexity on the order of  $\mathcal{O}(N_r^2)$, 
which mainly lies in the matrix-vector multiplication ${\bf{\Xi}}_t{{\bf{p}}_{i - 1}}$ per iteration. 
However, the calculation for $\zeta _t$ in \eqref{eq:LE_revised} is neglected in the above discussion,
which requires information on ${\hat{\bf{W}}}_t$.
According to \cite{ma_orthogonal_2017}, ${\tr}( {{{\hat{\bf{W}}}_t}\bar{\bf{G}}} )$ can be derived by 
$\sum_{i = 1}^{2N_r} {{\lambda _i}/({\lambda _i} + \frac{{\sigma_{\nu}^2}}{{2v_{t}^2}})} $, 
where $\lambda _i\;(i=1,\dots,2N_r)$ denotes the $i$-th eigenvalue of $\bar{\bf{G}}{\bar{\bf{G}}^T}$. 
These eigenvalues can be calculated only once outside the OAMP iterations because 
they remain constant during the iterations. 
Therefore, the computational cost is effectively reduced.

\subsection{Proposed CG-OAMP-NET}
\CheckRmv{
\begin{figure}[tbp]
	\centerline{\includegraphics[width=4.5in]{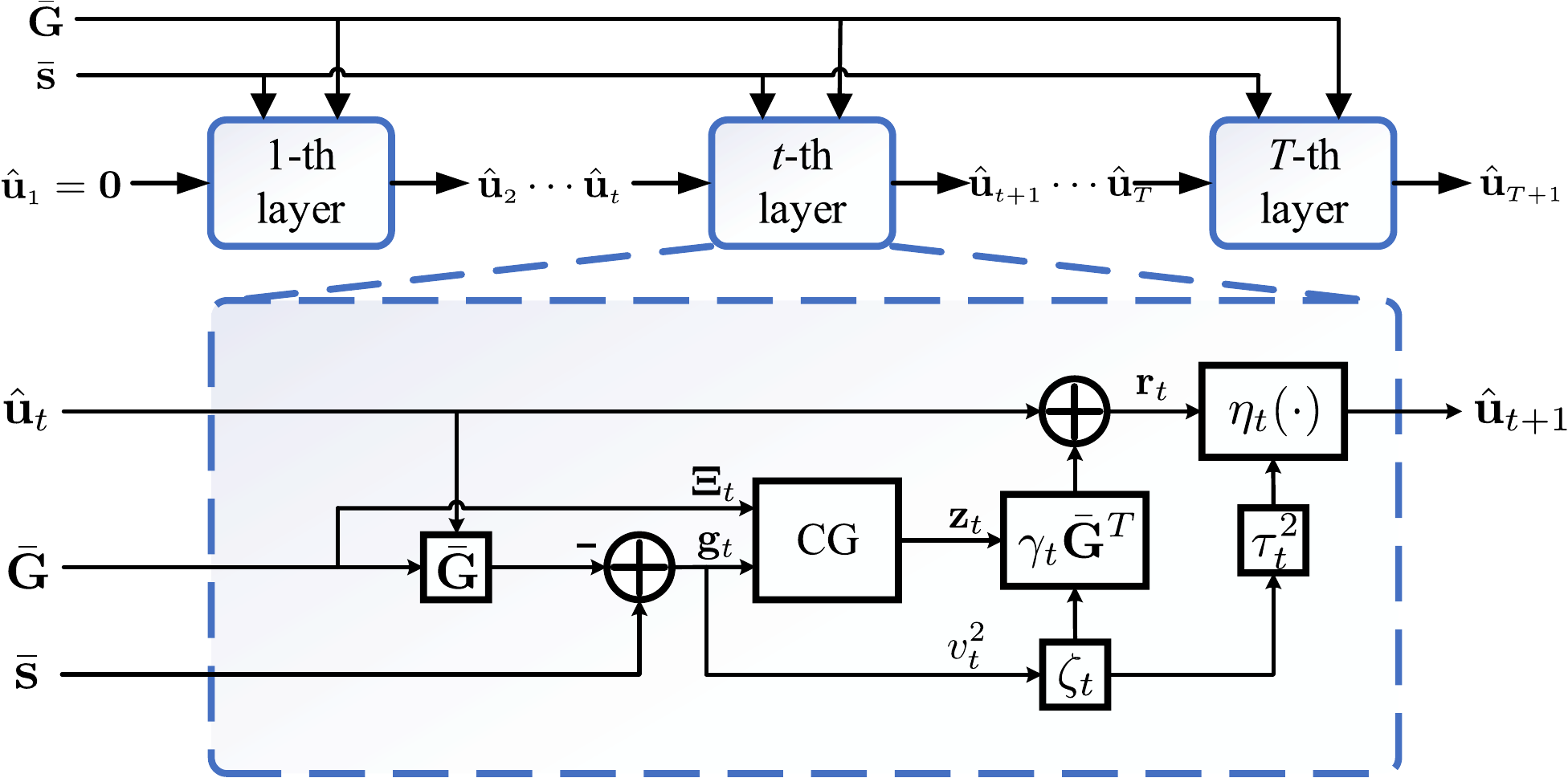}}
	\caption{Structure of the CG-OAMP-NET. 
	} 
	\label{fig:CG_OAMP_NET}
\end{figure}
}
\subsubsection{Network Architecture}
\figref{fig:CG_OAMP_NET} illustrates the structure of the $T$-layer CG-OAMP-NET.
The structure is derived by unfolding the iterations of CG-OAMP and adding some
layer-dependent tunable parameters to enhance detection performance. 
Each layer of the network can be separated into the following three parts: the preprocessing
module, the LE, and the NLE. We elaborate on these modules subsequently. 
The input of CG-OAMP-NET includes the channel matrix $\bar{\mathbf{G}}$ and the received signal $\bar{\mathbf{s}}$. 
In addition to $\bar{\mathbf{G}}$ and $\bar{\mathbf{s}}$,
the estimation $\hat{\bf{u}}_t$ generated by the ($t-1$)-th layer is another input
for the $t$-th layer of the network. 
The final output is the estimated symbol $\hat{\bf{u}}_{T+1}$.

\subsubsection{Preprocessing Module}
The preprocessing module includes a CG module for calculating the solution vector
$\mathbf{s}_t$ and an arithmetic unit for calculating the de-correlated coefficient $\zeta_t$. 
The CG module proceeds with an iteration as \algref{alg:CG} after deriving 
${\boldsymbol{\Xi}}_t$ and ${\mathbf{g}}_t$ in \eqref{eq:Xi} and \eqref{eq:yt}, respectively. 
We omit the details of this part in \figref{fig:CG_OAMP_NET} because it also presents an unfolded style. 
The output $\mathbf{z}_t$ and the de-correlated coefficient $\zeta_t$ are sent to the LE for the
following operations.

\subsubsection{Linear and Nonlinear Estimator}
LE and NLE, the cores of each layer, integrate four scalar trainable 
variables ${\Omega_t=\{\gamma_t,\theta_t,\phi_t,\xi_t\}}$. Thus, the OAMP stage 
is rewritten as\footnote{For the update of $v_t^2$, we still adopt \eqref{eq:v2} and set 
the minimum allowed variance $\epsilon$ and damping factor $\beta$ the same as those in CG-OAMP.}
\CheckRmv{
\begin{align}
	{{\bf{r}}_t} &= {\hat{\bf{u}}_t} + {\gamma _t}{\zeta _t}{\bar{\bf{G}}^T}{{\bf{z}}_t}, \label{eq:LE_NET}\\
	\tau _t^2 &= v_t^2( {\theta _t^2{\zeta _t} - 2{\theta _t} + 1} ), \label{eq:tau2_NET}\\
	{\hat{\bf{u}}_{t + 1}} &= {\eta _t}({{\bf{r}}_t},{\tau _t^2};{\phi _t},{\xi _t}). \label{eq:NLE_NET}
\end{align}
}
The LE in \eqref{eq:LE_NET} and the error variance estimator in \eqref{eq:tau2_NET} are
respectively modified from \eqref{eq:LE_revised} and \eqref{eq:tau2} by incorporating the parameters $(\gamma_t,\theta_t)$. 
The two parameters can be regarded as the update step sizes for ${\bf{r}}_t$ and $\tau_t^2$
\cite{ito_trainable_2019}. Equation \eqref{eq:MMSE} reveals that $ \mathbf{r}_t$ and $\tau_t^2$ are the prior
mean and variance of $\bar{\mathbf{u}}$, respectively, which are crucial for the accuracy of the MMSE estimate.
Hence, the parameters $(\gamma_t,\theta_t)$ can control the convergence behavior of the detector. 
In addition, the expression for $\tau_t^2$ in \eqref{eq:tau2_NET} originates from the variance estimator in 
\cite{he_model-driven_2020}, and the derivation is similar to that for \eqref{eq:tau2}, which is also 
provided in the Appendix.

The NLE $\eta_{t}(\cdot)$ in \eqref{eq:NLE_NET} can be represented by
\CheckRmv{
\begin{equation}
	\eta_{t}({\bf{r}}_t,\tau_{t}^2;\phi_t,\xi_t)
	=\phi_t({\bf{E}}\{\bar{\bf{u}}|{\bf{r}}_t,\tau_t\}-\xi_{t}{\bf{r}}_t), \label{eq:NLE_NET_df}
\end{equation}
}
which combines the contribution of the MMSE denoiser in \eqref{eq:MMSE} and the LE.
A divergence-free $\eta_{t}$ can be constructed without relying on the prior information
by tuning the parameters $(\phi_t,\xi_t)$ in \eqref{eq:NLE_NET_df} via DL \cite{he_model-driven_2020}. 
This divergence-free function is critical for preserving orthogonality between the two local estimators.

The total number of trainable variables in the network is only $4T$,
which is independent of the system size. 
Thus, the training process of CG-OAMP-NET will be simple and stable. 

\subsubsection{Relation to Other Detectors} 
First, we elaborate on the relationship between CG-OAMP and CG-OAMP-NET.
The main difference between the two detectors lies in the trainable variables $\{\Omega_t\}_{t=1}^T$. 
In other words, the CG-OAMP is a special case of CG-OAMP-NET with $\gamma_t=\theta_t=\phi_t=1$ and $\xi_t=0$.
The introduction of these variables originates from the difficulty in achieving the
prerequisites of conventional AMP-type algorithms (AMP, OAMP, and CG-OAMP) in 
realistic finite-dimensional systems. 
For example, these AMP-type algorithms {severely deteriorate} when the system size is small,
or the channel has strong correlations. 
However, {CG-OAMP-NET can tune the variables to adapt to realistic environments through the DL.}
Thus, the network can rapidly converge, reach effective performance, and demonstrate adequate
flexibility to handle various channel environments.
This idea is similar to that of OAMP-NET in \cite{he_model-driven_2020}. 
However, the proposed method is different from OAMP-NET in that the network contains a
CG module, and the OAMP stage is free of matrix inversion, which renders the structure efficient.

\subsection{Application to the CP-Free Scenario} \label{sec:CP_free}

\CheckRmv{
\begin{figure*}[!t]
\setlength{\abovecaptionskip}{-0.2cm}
\setlength{\belowcaptionskip}{-0.0cm}
	\centerline{\includegraphics[width=3.85in]{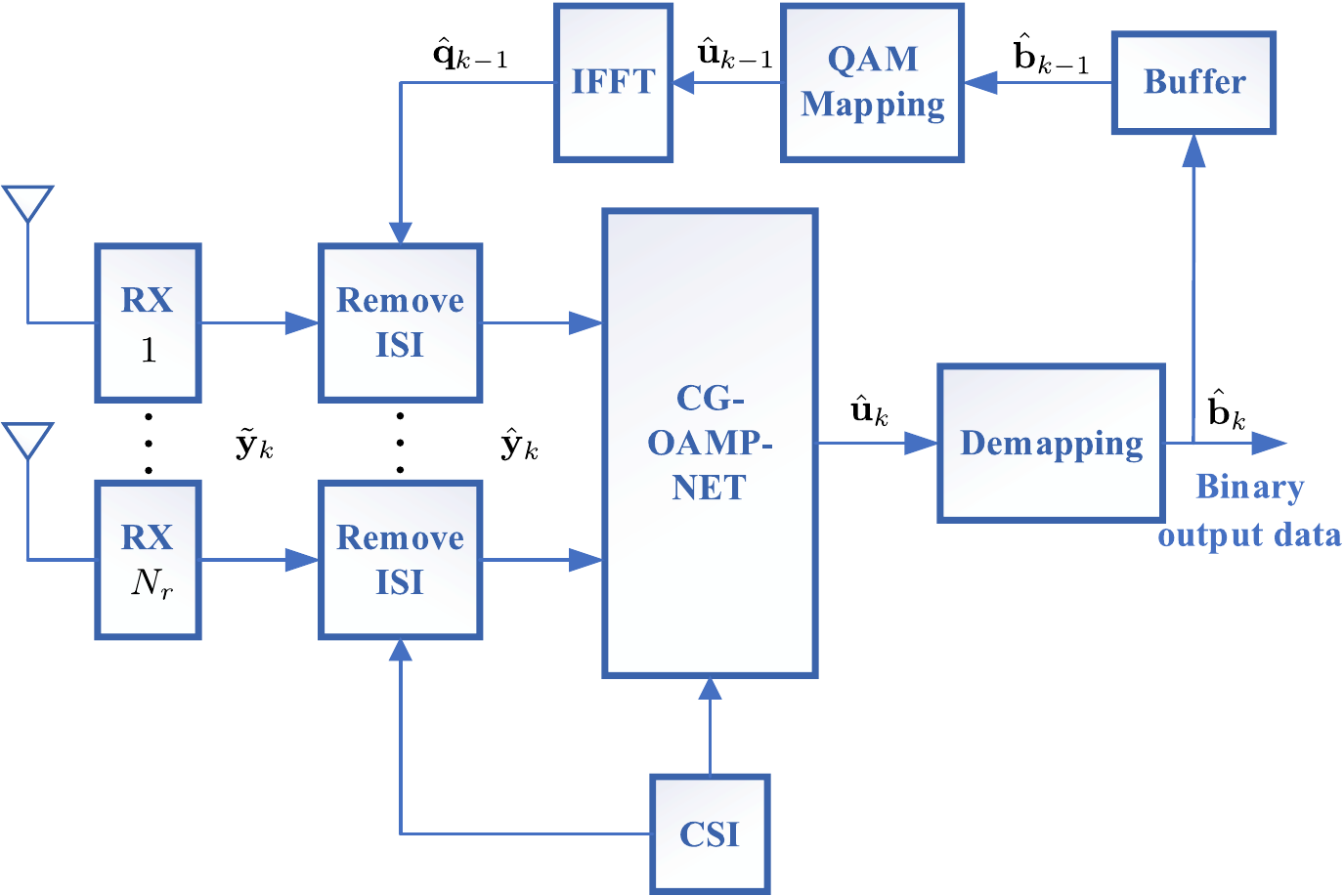}}
	\caption{Block diagram of an uncoded CP-free MIMO-OFDM system with  
	a model-driven MIMO-OFDM receiver.
	The transmitter is omitted herein because it has the same structure 
	as that in \figref{fig:MIMO_OFDM_system} except that no CP is inserted.}
	\label{fig:CP_free_model}
\end{figure*}
}

Different from the traditional MIMO-OFDM system discussed in Section \ref{sec:sys_model}, 
we then consider a CP-free one shown in \figref{fig:CP_free_model},
that is, no CP is appended to the channel input. By defining
\[\tilde{\mathbf{u}}_k = {\left[ \mathbf{u}_{k,{0}}^T,\dots, \mathbf{u}_{k,n}^T, \dots,
\mathbf{u}_{k,{N_c} - 1}^T \right]^T}
\in {\mathbb{C}^{N_c{N_t} \times 1}}, \]
the $k$-th received signal $\tilde{\mathbf{y}}_k$, {\Tx} signal $\tilde{\mathbf{q}}_k$, 
and noise $\tilde {\mathbf{w}}_k$ in the time domain are given by 
\CheckRmv{
\begin{align}
    \tilde{\mathbf{y}}_k &= \tilde{\mathbf{H}}_k \tilde{\mathbf{q}}_k - 
    {\mathbf{A}_k \tilde{\mathbf{q}}_k} + 
    {\mathbf{A}_{k-1}\tilde{\mathbf{q}}_{k-1}} +
	\tilde{\mathbf{w}}_k \nonumber \\
	&=(\tilde{\mathbf{H}}_k-\mathbf{A}_k) \tilde{\mathbf{F}}^{\mathrm{H}} \tilde{\mathbf{u}}_k+\mathbf{A}_{k-1} 
	\tilde{\mathbf{q}}_{k-1}+\tilde{\mathbf{w}}_k, 
    \label{eq:Rx_CP_free}
\end{align}
}
\[\tilde{\mathbf{q}}_k = {\left[ \mathbf{q}(kN_c+0)^T, \dots ,\mathbf{q}(kN_c+N_c-1)^T \right]^T}
\in {\mathbb{C}^{N_c{N_t} \times 1}}, \] 
\[\tilde {\mathbf{w}}_k = {\left[ \mathbf{w}(kN_c+0)^T, \dots ,\mathbf{w}(kN_c+N_c-1)^T \right]^T}
\in {\mathbb{C}^{N_c{N_r} \times 1}}. \] 
The second and third terms of the first equality in \eqref{eq:Rx_CP_free} denote the ICI and ISI, respectively. 
Meanwhile, $\tilde{\mathbf{q}}_k$ and $\tilde{\mathbf{q}}_{k-1}$ denote the $k$-th and $(k-1)$-th 
transmitted MIMO-OFDM signal vectors, respectively.
The second equality of \eqref{eq:Rx_CP_free} originates from the relation between the frequency and time domains, that is, 
$\tilde{\mathbf{q}}_k=\tilde{\mathbf{F}}^{\mathrm{H}}\tilde{\mathbf{u}}_k=(\mathbf{F}^{\mathrm{H}} \otimes \mathbf{I}_{N_t}) \tilde{\mathbf{u}}_k$,
where $\mathbf{F}$ is an FFT matrix with $N_c \times N_c$ dimension, 
of which the $(m,n)$-th element is equal to $\exp (-j2\pi(mn/N_c))/\sqrt{N_c}$. 
The multipath channel matrix $\tilde{\mathbf{H}}_k$ is an $N_c N_r \times N_c N_t$ blocked-circulant
matrix 
with the first block of columns being 
$\left[\mathbf{H}_{k,0}^T, \dots ,\mathbf{H}_{k,L-1}^T, \mathbf{0}, \dots ,\mathbf{0}\right]^{T}
\in {\mathbb{C}^{N_c{N_r} \times N_t}}$ \cite{van_zelst_implementation_2004}.
The interference channel matrix $\mathbf{A}_k$ is also an $N_c N_r \times N_c N_t$ blocked matrix
and given by the following:
\[\mathbf{A}_k=\left[
\begin{array}{ccccccc}
	\mathbf{0} & \cdots & \mathbf{0} & \mathbf{H}_{k,L-1} & \cdots & \cdots & \mathbf{H}_{k,1} \\
	\mathbf{0} & \cdots & \mathbf{0} & \mathbf{0} & \mathbf{H}_{k,L-1} & \cdots & \mathbf{H}_{k,2} \\
	\vdots & \cdots & \vdots & \ddots & \ddots & \ddots & \vdots \\
	\mathbf{0} & \cdots & \mathbf{0} & \ddots & \ddots & \mathbf{0} & \mathbf{H}_{k,L-1} \\
	\vdots & \cdots & \vdots & \ddots & \ddots & \ddots & \vdots \\
	\mathbf{0} & \cdots & \mathbf{0} & \mathbf{0} & \ddots & \cdots & \mathbf{0}
\end{array}
\right].\]

$\mathbf{A}_k$ and $\mathbf{A}_{k-1}$ are reduced to $\mathbf{0}$ in the aforementioned 
MIMO-OFDM system with sufficient CP, and ICI and ISI are absent.
However, the interference terms in \eqref{eq:Rx_CP_free} complicate signal detection.
From \figref{fig:CP_free_model}, we develop a model-driven MIMO-OFDM receiver for this CP-free scenario,
which includes a CG-OAMP-NET for data detection, 
a QAM demapping module, and other modules for residual ISI elimination. 
Eliminating the residual ISI at the receiver before the detection process is necessary.
As shown in \figref{fig:CP_free_model}, the recovered bit stream $\hat{\mathbf{b}}_{k-1}$
from the $(k-1)$-th detection process is stored in the buffer module for the $k$-th detection.
The estimated signals in the frequency and time domains,
which are respectively denoted by $\hat{\mathbf{u}}_{k-1}$ and $\hat{\mathbf{q}}_{k-1}$, 
are generated successively from $\hat{\mathbf{b}}_{k-1}$ as feedback after QAM mapping and IFFT. 
Thus, the ISI can be removed from the original received signal $\tilde{\mathbf{y}}_k$ as follows\footnote{{In the coded transmission, we can use error detection codes (e.g., cyclic redundancy checks) to identify the number of error bits in the previous detection \cite{shlezinger_viterbinet_2020,teng_syndrome-enabled_2020} and avoid the wrong ISI cancellation that may worsen the detection performance.}}: 
\CheckRmv{ 
\begin{align}
	\hat{\mathbf{y}}_k &=\tilde{\mathbf{y}}_k-
	\mathbf{A}_{k-1} \hat{\mathbf{q}}_{k-1} \nonumber  \\
	&=(\tilde{\mathbf{H}}_k-\mathbf{A}_k) \tilde{\mathbf{F}}^{\mathrm{H}} 
	\tilde{\mathbf{u}}_k+\mathbf{A}_{k-1} 
	\tilde{\mathbf{q}}_{k-1}+\tilde{\mathbf{w}}_k-\mathbf{A}_{k-1}
	\hat{\mathbf{q}}_{k-1} \nonumber \\
	&\approx \mathbf{C}_k \tilde{\mathbf{u}}_k+\tilde{\mathbf{w}}_k, 
\label{eq:remove_isi}
\end{align} 
}
where the matrix $\mathbf{C}_k=(\tilde{\mathbf{H}}_k-\mathbf{A}_k) \tilde{\mathbf
{F}}^{\mathrm{H}}$ can be obtained by the known CSI. 
Equation \eqref{eq:remove_isi} is a signal recovery problem similar to \eqref{eq:MIMO_model}.
Thus, we convert the  ${M \times Q}$ complex-valued system \eqref{eq:remove_isi} with $M=N_cN_r$ and $Q=N_cN_t$
into an equivalent ${{2}M \times {2}Q}$ real-valued system as follows: 
\CheckRmv{
\begin{equation}
	{\mathbf{y}}_r = {\mathbf{C}}_r{\mathbf{u}}_r + {\mathbf{w}}_r,
\label{eq:real_domain_system}
\end{equation}
}
where $\mathbf{y}_r \in \mathbb{R}^{2M\times 1}$, $\mathbf{u}_r\in \mathbb{R}^{2Q\times 1}$, 
$\mathbf{w}_r \in \mathbb{R}^{2M\times 1}$, and $\mathbf{C}_r \in \mathbb{R}^{2M\times 2Q} $ are the 
real-valued forms of $\hat{\mathbf{y}}_k$, $\tilde{\mathbf{u}}_k$, $\tilde{\mathbf{w}}_k$,
and $\mathbf{C}_k$, respectively.
Then, \eqref{eq:real_domain_system} can be solved by the proposed CG-OAMP-NET based on ${\mathbf{y}}_r$, 
${\mathbf{C}}_r$, and the noise variance $\sigma_w^2$.
The details are organized in \algref{alg:CG_OAMP}. 
\renewcommand{\algorithmicrequire}{\textbf{Input:}}
\CheckRmv{
\begin{algorithm}[!hb]
	\caption{CG-OAMP-NET for CP-free MIMO-OFDM}
	\label{alg:CG_OAMP}
	\begin{algorithmic} 
		\REQUIRE Received signal ${\bf{y}}_r$, channel matrix ${\bf{C}}_r$, noise variance $\sigma_w^2$.
		\STATE \textbf{Initialize:} ${\hat{\bf{{u}}}_1} = {\bf{0}}$, ${v_1^2} = 1$,
		calculate the eigenvalues $\lambda _i\;(i=1,\dots,2M)$ of ${\bf{C}}_r{\bf{C}}_r^T$.
		\FOR{$t=1$ to $T$}
			\STATE 1: Derive ${\bf{\Xi }}_t$ and ${{\bf{g}}}_t$, and solve the linear 
			system ${\bf{\Xi }}_t {{\bf{z}}}_t = {{\bf{g}}}_t$ by CG:
			\begin{align}
				{\boldsymbol{\Xi }}_t &= {\mathbf{C}}_r {\mathbf{C}}_r^T + \frac{\sigma _w^2}{2v_{t}^2}{\mathbf{I}}, \label{eq:Xi_cpfree}\\
				{\mathbf{g}}_t &= {\mathbf{y}}_r - {\mathbf{C}}_r \hat{\mathbf{u}}_t, \label{eq:yt_cpfree}\\
				{{\bf{z}}_t} &= \textrm{CG}\{ {\bf{\Xi}}_t,{\bf{g}}_t \}. \label{eq:CG}
			\end{align} \\
			 2: Calculate the de-correlated coefficient $\zeta _t$:
			\begin{equation}
				\frac{1}{\zeta _t} = \frac{1}{2Q} {\sum\limits_{i = 1}^{2M} {{\lambda _i}/({\lambda _i} + 
				\frac{{\sigma _w^2}}{{2v_{t}^2}})}}. \label{eq:nor_coef}
			\end{equation} \\
			3: The OAMP stage: 
			\begin{gather}
				{{\bf{r}}_t} = {\hat{\bf{u}}_t} + {\gamma _t}{\zeta _t}{{\bf{C}}_r^T}{{\bf{z}}_t},\\
				\tau _t^2 = v_t^2( {\theta _t^2{\zeta _t} - 2{\theta _t} + 1} ),\\
				{\hat{\bf{u}}_{t + 1}} = {\eta _t}({{\bf{r}}_t},{\tau _t^2};{\phi _t},{\xi _t})
				=\phi_t({\bf{E}}\{{\bf{u}}_r|{\bf{r}}_t,\tau_t\}-\xi_{t}{\bf{r}}_t), \\
				v_{t+1}^2 = \frac{ {\| {\bf{y}}_r - {\bf{C}}_r \hat{\bf{u}}_{t+1}\|}^2 - M\sigma _w^2}
				{{\tr}( {\bf{C}}_r^T {\bf{C}}_r )}.
			\end{gather}
		\ENDFOR
		\ENSURE Estimated symbol ${\hat{\bf{{u}}}_{T+1}}$.
	\end{algorithmic} 
\end{algorithm}
}

\subsection{Complexity Analysis}  \label{sec:complexity}
\CheckRmv{
\begin{table*}[tbp]
	\caption{Complexity and running time comparison of different algorithms for CP-free MIMO-OFDM}
	\begin{center}
	\begin{tabular}{lccc}
		\toprule
		Algorithms & Computational complexity & \multicolumn{2}{c}{Running time$^{\mathrm{a}}$} \\
		& &  ${\text{8}\times\text{8}}$  & ${\text{32}\times\text{32}}$  \\
		\midrule
		OAMP/OAMP-NET & ${\cal O}
		\Big((M^3+M^2+MQ)T_{\OAMP}\Big)$ & 0.506 & 18.081\\
		MAMP & ${\cal O}\Big(MQT_{\MAMP}+(M+Q)T_{\MAMP}^2+
		T_{\MAMP}^3+L_d^3T_{\MAMP}\Big)$  &  0.192 & \textbf{1.592} \\
		CG-OAMP/CG-OAMP-NET & ${\cal O}\Big(({I_{{\rm{CG}}}}
		(M^2 + {M}) + M^2 + {M}{Q})T_{\CGOAMP}\Big)$  & \textbf{0.140} & 2.572\\
		\bottomrule
		\multicolumn{4}{l}{$^{\mathrm{a}}$The running time (in Seconds) for the complete 
		detection process of one MIMO-OFDM symbol when $N_c=64$,}\\
		\multicolumn{4}{l}{$T_{\OAMP}=T_{\CGOAMP}=5$,
		$T_{\MAMP}=30$, $I_{\textrm{CG}}=50$, and $L_d=3$. {The number of iterations is chosen according to the }}\\ 
		\multicolumn{4}{l}{{convergence evaluation in \figref{fig:cp_free_convergence} of Section \ref{sec:convergence}.}
		The antenna configuration is either ${\text{8}\times\text{8}}$ or ${\text{32}\times\text{32}}$, resulting}\\
		\multicolumn{4}{l}{
			in $M=Q=512\;\text{or}\;2048$. Modulation scheme is 16QAM. All the algorithms are implemented 
		on the same PC}\\ 
		\multicolumn{4}{l}{with an Intel Core i5-7300 CPU @ 2.50 GHz and 16 GB memory.}
	\end{tabular}
	\label{tab:complexity}
	\end{center}
\end{table*}
}
We compare the computational complexity of different algorithms for CP-free MIMO-OFDM
detection, including OAMP/OAMP-NET, CG-OAMP/CG-OAMP-NET, and MAMP \cite{liu_memory_2021}.
The number of iterations (layers) in the algorithms is $T_{\OAMP}$, $T_{\CGOAMP}$, and $T_{\MAMP}$.
MAMP, a low-cost AMP-type algorithm, is {\BO} for large-scale systems with unitarily-invariant channel matrices.
MAMP needs some parameters related to the eigenvalue distribution $P_{\lambda}$ of 
${\mathbf{C}}_r{\mathbf{C}}_r^T$ in the iterative process \cite{liu_memory_2021}. Thus,  
we assume that $P_{\lambda}$ is known in this comparison for ease of analysis.
\tabref{tab:complexity} presents the complexity of different detectors,
where the result of MAMP is from \cite{liu_memory_2021}, and $L_d$ is the damping length in MAMP.
The LMMSE estimation is dominant in the per-iteration cost of OAMP,
which comprises a matrix inversion and three matrix-vector multiplications,
and results in a cost of ${\cal O}(M^3+M^2+MQ)$.
CG-OAMP avoids the matrix inversion by using inner CG iterations, 
and the complexity per CG iteration, as shown in \algref{alg:CG},  
is dominated by the computation of ${\bf{\Xi }}_t{{\bf{p}}_{i - 1}}$ 
and three vector multiplications. 
Hence, the complete CG process \eqref{eq:CG} for each iteration of CG-OAMP costs 
${\cal O}({I_{{\rm{CG}}}}(M^2 + {M}))$ in place of ${\cal O}(M^3)$ due to matrix inversion. 
The de-correlated coefficient $\zeta_t$ can also be derived from $P_{\lambda}$ \cite{takeuchi_rigorous_2017},
and the cost is ignored herein because the counterpart is ignored in \cite{liu_memory_2021}.
Moreover, OAMP-NET and CG-OAMP-NET exhibit the same complexity as their prototypes 
because the required operations do not change.
We can conclude from \tabref{tab:complexity} that MAMP and the proposed CG-OAMP-NET are more efficient compared with OAMP/OAMP-NET due to the substantially low per-iteration cost.

Moving beyond the rough analysis in ${\cal O}(\cdot)$, we also investigate a running time
test for the detection process of one MIMO-OFDM symbol. 
The system parameters are shown below in \tabref{tab:complexity}.
Compared with the running time of OAMP/OAMP-NET, MAMP and CG-OAMP-NET need approximately 2.64 and
3.61 times less in an ${\text{8}\times\text{8}}$ MIMO system and 11.36 and 7.03 times less in a ${\text{32}\times\text{32}}$ MIMO system, respectively. 
{MAMP is inferior to CG-OAMP-NET in terms of running time in the relatively small system because the former needs 30 iterations to converge, while the latter requires only five layers (\figref{fig:cp_free_convergence} of Section \ref{sec:convergence}). However, MAMP has a lower per-iteration cost with respect to the system size compared with CG-OAMP-NET because the cost mainly lies in the matrix-vector multiplications with a complexity of only ${\cal O}(MQ)$. Thus, MAMP has an advantage when the system size increases, and the iteration times are fixed.} 

\section{Simulation Results}    \label{sec:simu_results}
The performance of CG-OAMP-NET is evaluated in this section by simulations and compared
with some state-of-the-art schemes,
including the MAMP and the OAMP-NET as the baseline.
First, we provide the implementation details. 
Then, the convergence property and detection performance of CG-OAMP-NET are shown. 
Finally, we compare the performance and spectral efficiency of the MIMO-OFDM  system with sufficient CP (SCP)
and the CP-free scheme developed in Section \ref{sec:CP_free} {and investigate the impact of imperfect CSI on CG-OAMP-NET.}

\vspace{-0.5cm}
\subsection{Implementation Details}
We investigate a MIMO-OFDM system with $N_c=64$ subcarriers and a 16-tap channel.
The CP length $N_g$ equals the channel delay spread in SCP systems, while no
CP is added for CP-free systems. 
No channel coding is also used in our simulation.
The simulated signal-to-noise ratio (SNR) at the receiver is defined as follows:
\CheckRmv{
\begin{equation}
	{\textrm{SNR}} = 10{\log_{10}}\Big(\frac{ {\bf{E}}\{ {\left\| \bar{{\mathbf{G}}} \bar{{\mathbf{u}}} 
	\right\|}^2 \} } { {\bf{E}} \{ {\left\|\bar{\boldsymbol{\nu}}\right\|}^2 \} }\Big).
\end{equation}
}

The CG-OAMP-NET is implemented with TensorFlow. We train the network for 1000 epochs.
The training and validation sets for SCP systems comprise 5000 and 10,000 different 
samples for each epoch. 
Each sample is a randomly generated pair 
$\{\bar{{\mathbf{u}}}^{(i)},\bar{\mathbf{s}}^{(i)},\bar{{\mathbf{G}}}^{(i)}\}$, 
where the transmitted data $\bar{\mathbf{u}}^{(i)}$ is the label, and the received signal
$\bar{\mathbf{s}}^{(i)}$ together with the channel matrix $\bar{\mathbf{G}}^{(i)}$ are the features.
In our simulation, $\bar{\mathbf{G}}^{(i)}$ is drawn from specific channel models, including 
the Rayleigh fading channel, the correlated MIMO channel, and the wireless world initiative 
for new radio model (WINNER II) \cite{kyosti_ist-4-027756_2007}. {We use the WINNER II channel under typical urban scenarios, non-line-of-sight case, where the carrier frequency is 2.6 GHz, and the max delay spread is 16 samples. The antenna array configurations are uniform linear arrays with 10 wavelengths of antenna spacing to reduce spatial correlations.}
We conduct mini-batch training with stochastic gradient descent in the training stage,
where the batch size is chosen as 100. 
The optimizer is chosen as Adam 
with an initial learning rate of 0.001. The loss function is given by $l_2$ loss\footnote{{We choose the $l_2$ loss function for easy comparison with the results in \cite{he_model-driven_2020}. We also train the CG-OAMP-NET with the binary cross entropy \cite{cammerer_trainable_2020} and do not find performance improvement compared with the $l_2$ loss.}},
\CheckRmv{
\begin{equation}
	l_2 = \frac{1}{S}\sum\limits_{i=1}^{S}{\left\| { \hat{\mathbf{u}}_{T + 1}^{(i)}}
	(\bar{\mathbf{s}}^{(i)},\bar{\mathbf{G}}^{(i)}) - \bar{\mathbf{u}}^{(i)} \right\|} ^2,
\end{equation}
} 
where ${ \hat{\mathbf{u}}_{T + 1}^{(i)}}$ is the output prediction of the network, and $S$ is the batch size. 
The training and validation sets for CP-free systems are respectively resized to 500 and 100 samples, 
with each sample replaced by $\{{\mathbf{u}}_r^{(i)},{\mathbf{y}}_r^{(i)},{\mathbf{C}}_r^{(i)}\}$.
We generate data in the test stage to evaluate the network until the number of bit errors exceeds 1000.
Except for special instructions, we train and test the network under the same settings,
including SNR, modulation scheme, and channel model.
Moreover, we set the damping length in MAMP as $L_d=3$ in all the simulations. 

\vspace{-0.5cm}
\subsection{Convergence Property} \label{sec:convergence}
\CheckRmv{
\begin{figure*}[tbp]
	\centering
	\subfigure[${\text{128}\times\text{128}}$ MIMO, \SNR{=}{12}]{
		\label{fig:QAM_12dB_128MIMO}
		\includegraphics[width=.31\textwidth]{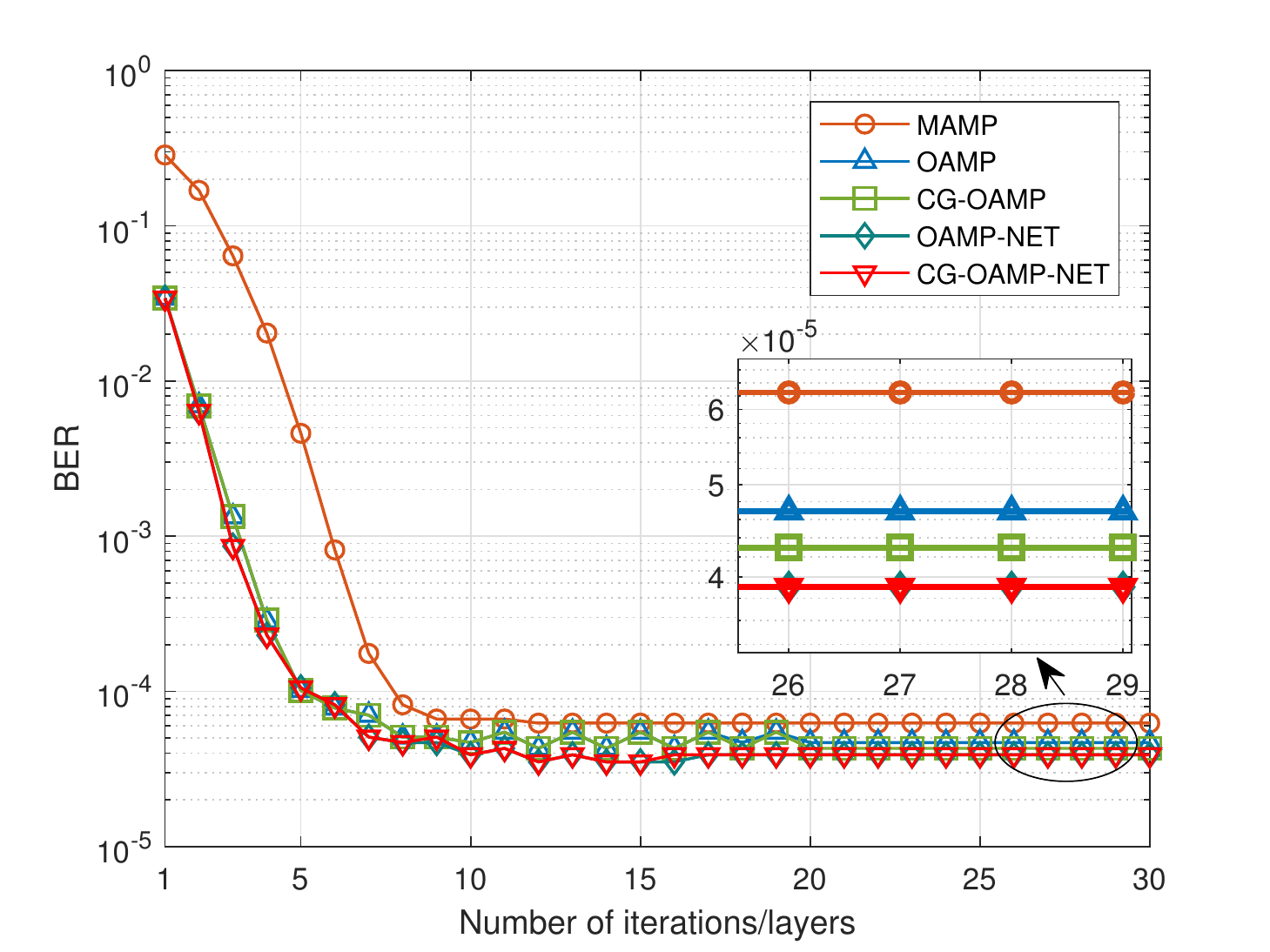}
	}
	\subfigure[${\text{8}\times\text{8}}$ MIMO, \SNR{=}{12}]{
		\label{fig:QAM_12dB_8MIMO}
		\includegraphics[width=.31\textwidth]{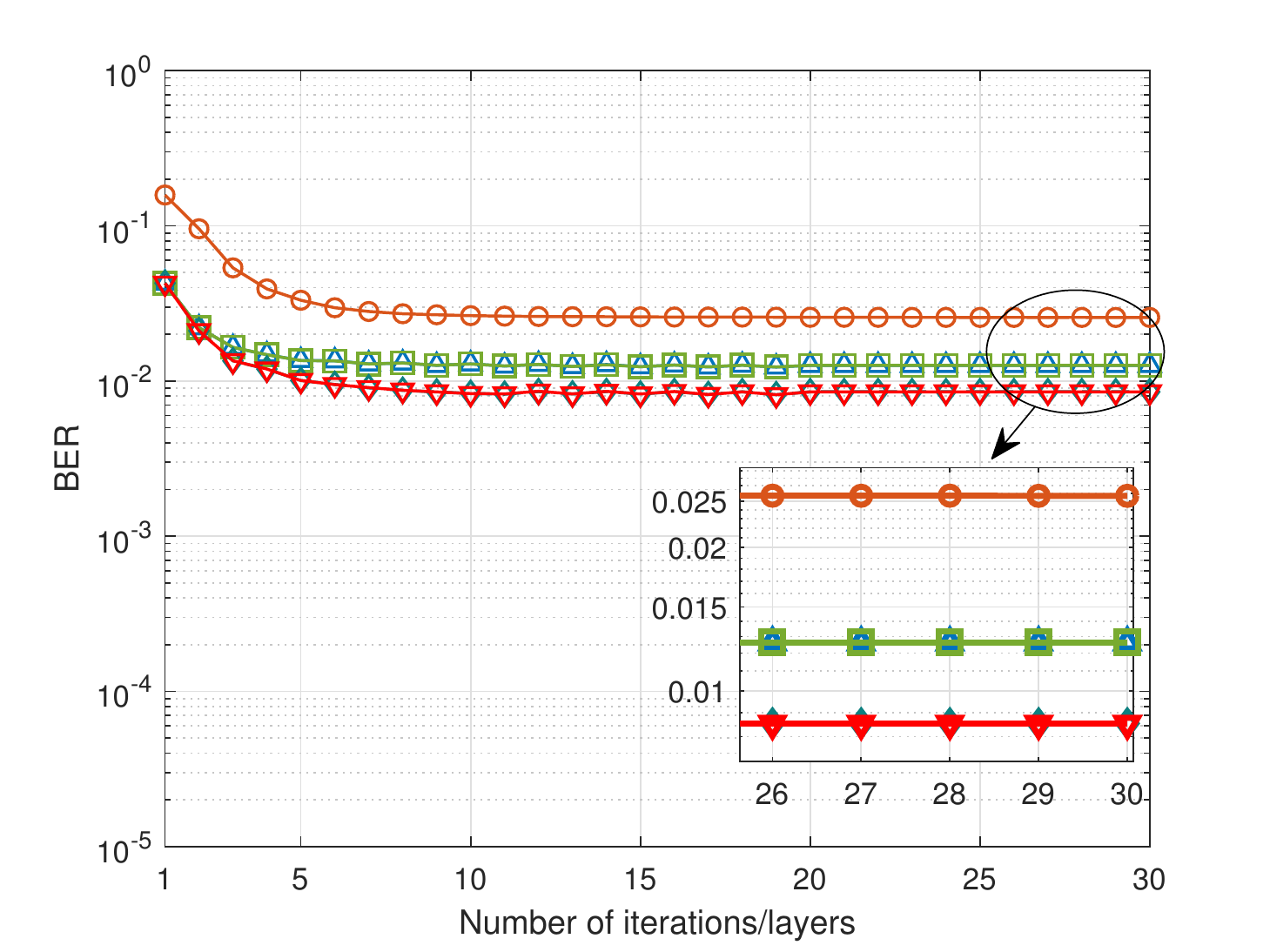}
	}
	\subfigure[${\text{8}\times\text{8}}$ MIMO, \SNR{=}{25}]{
		\label{fig:QAM_25dB_8MIMO}
		\includegraphics[width=.31\textwidth]{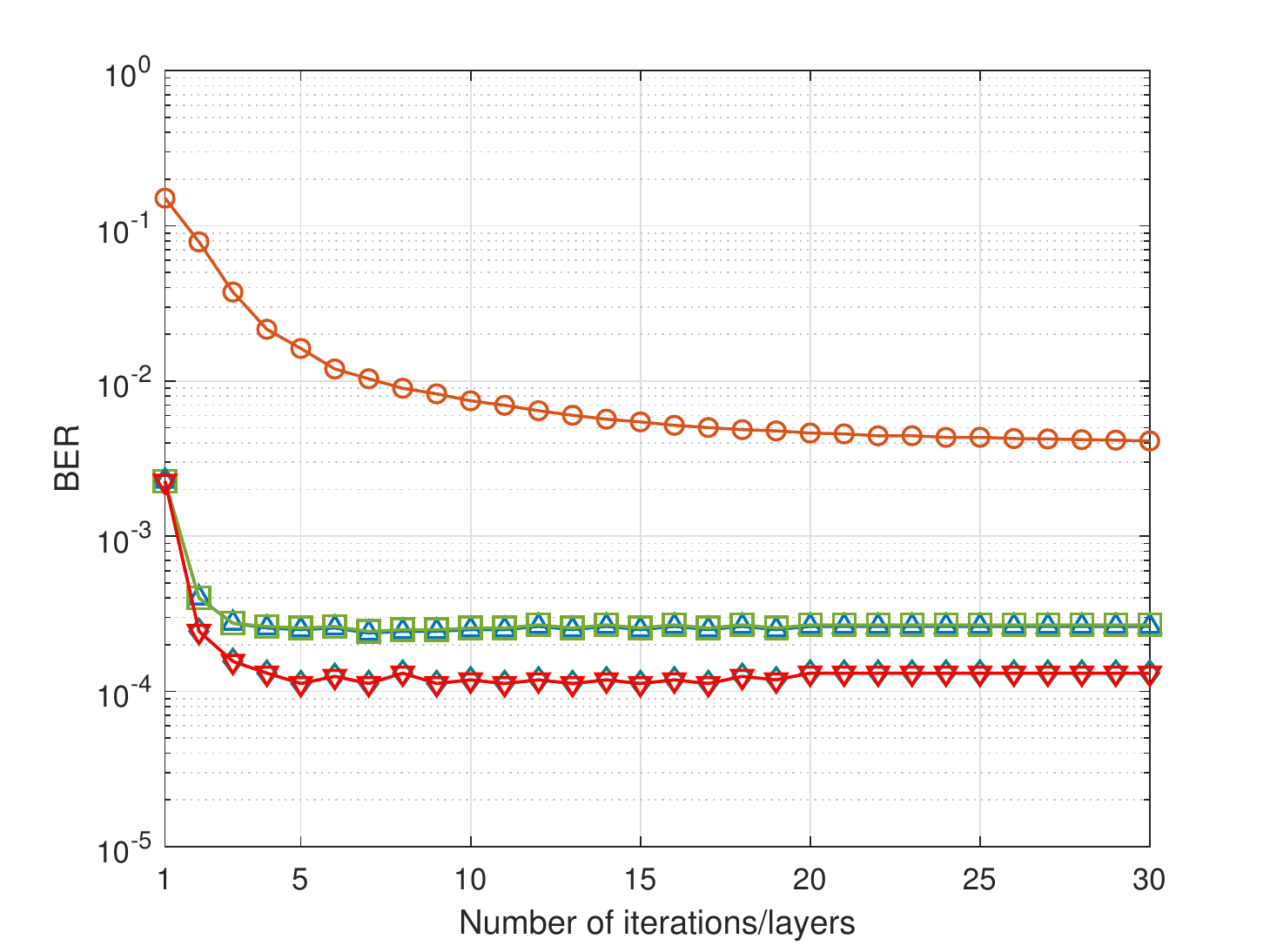}
	}
	\caption{{ BER versus the number of iterations (layers) under QPSK modulation.} The results are averaged over {1000} independent trails. The number of CG iterations in CG-OAMP/CG-OAMP-NET is $I_{\text{CG}}=50$.}
	\label{fig:QAM}
\end{figure*}
}

{We demonstrate the convergence property of the competing methods in this subsection. 
 \figref{fig:QAM} illustrates the BER performance of the proposed algorithm versus the number of iterations (layers) under QPSK modulation
  with IID Rayleigh MIMO channels. Each element $g_{i,j}$ of the channel matrix $\mathbf{G}$ satisfies $g_{i,j} \sim \mathcal{CN}(0,1/N_r)$. Figs.~\ref{fig:QAM_12dB_128MIMO} and \ref{fig:QAM_12dB_8MIMO} show the results under \SNR{=}{12} with different system dimensions. The figures demonstrate that MAMP can have a similar performance to OAMP only for a relatively high-dimensional MIMO (i.e., ${\text{128}\times \text{128}}$ MIMO) system. However, MAMP has serious performance degradation and cannot converge to OAMP anymore in a small-sized \Times{8}{8} MIMO system. By contrast, OAMP-NET can achieve a lower BER than its prototypical methods in both settings. This improvement can be attributed to the use of DL, which tunes the key parameters in the network to overcome the degradation of the prototypical algorithms. Furthermore, the BER curves of the CG-based revisions and those of the versions using direct matrix inverse for OAMP and OAMP-NET exactly match, suggesting that the introduction of CG does not incur any performance loss. \figref{fig:QAM_25dB_8MIMO} shows the BER results at the high SNR regime. In this situation, MAMP needs additional iterations for convergence, while the counterparts of the other algorithms do not substantially change.  Therefore, we set OAMP, CG-OAMP, and the corresponding NNs to five layers (${T=5}$) in the sequel except for special instructions. The number of iterations in MAMP is 15 when \SNR{<}{25} and increases to 30 for \SNR{=}{25} or above. The number of inner CG iterations in CG-OAMP and CG-OAMP-NET is set as 50 ($I_{\text{CG}}=50$) to adapt to a wide range of SNRs and different channels.}

\CheckRmv{
\begin{figure}[tbp]
	\centering
	\subfigure[$\kappa=100$]{
		\label{fig:cp_free_convergence.sub.1}
		\includegraphics[width=3.0in]{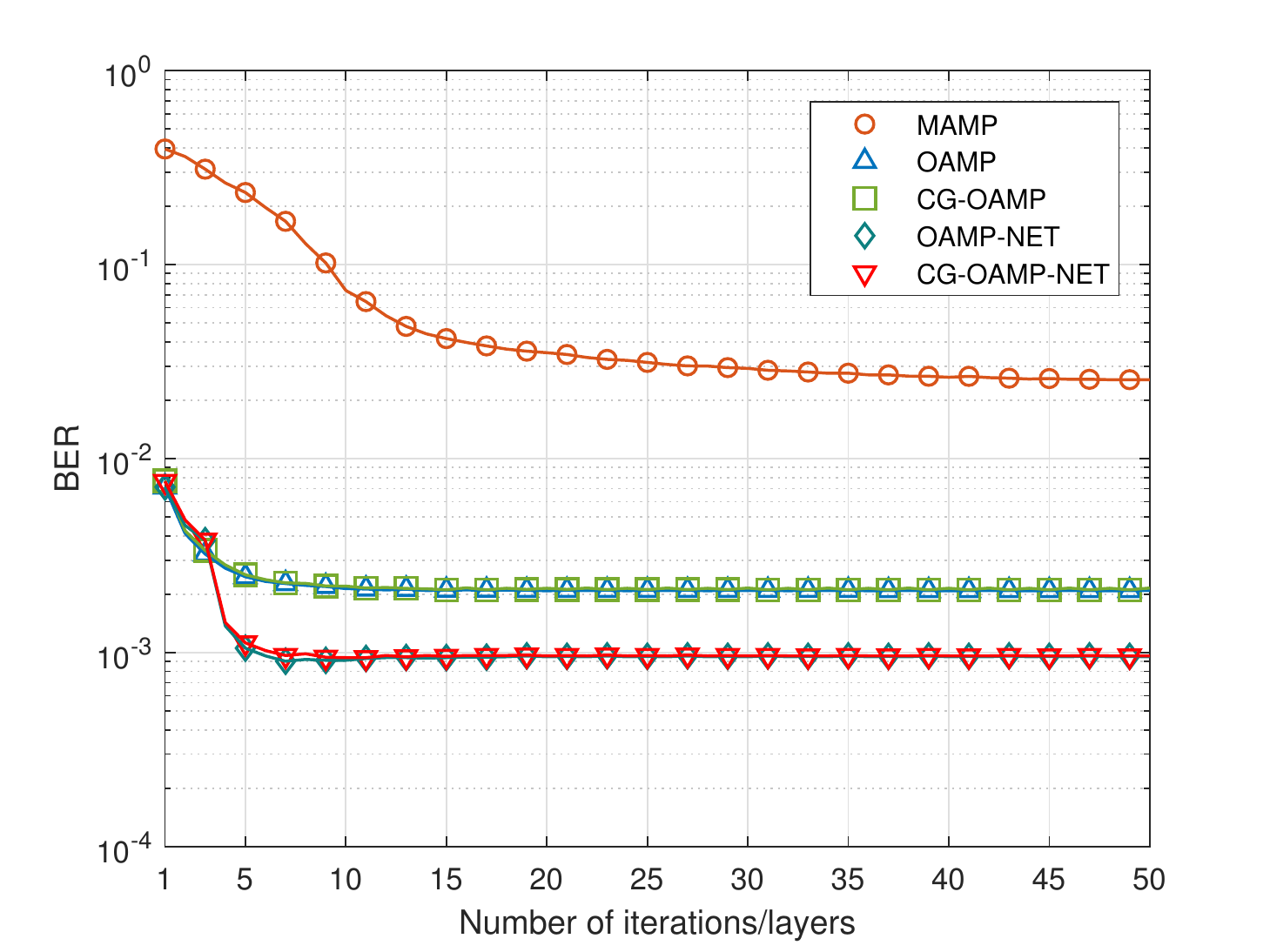}
	}
	\subfigure[$\kappa=200$]{
		\label{fig:cp_free_convergence.sub.2}
		\includegraphics[width=3.0in]{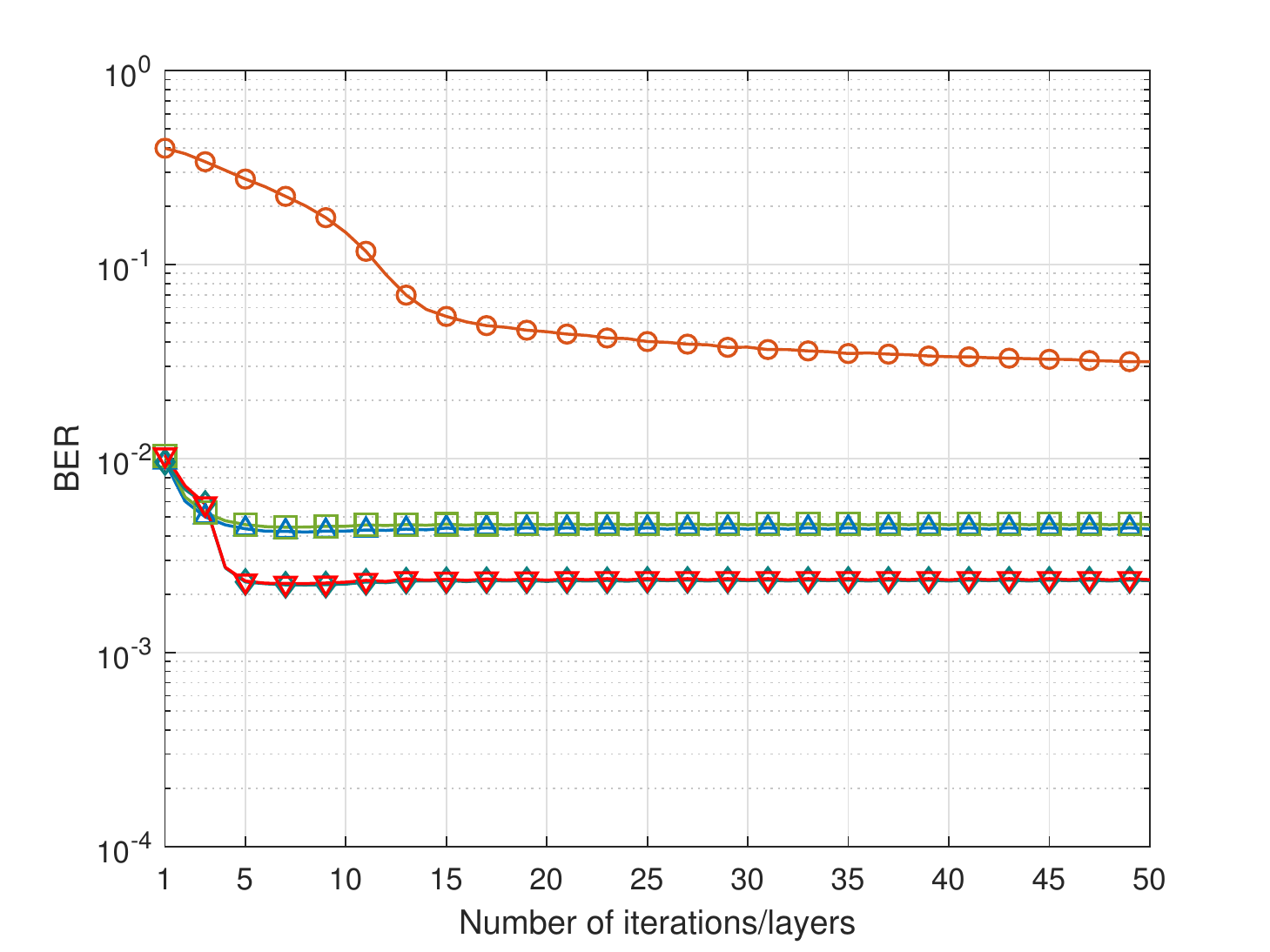}
	}
	\setlength{\abovecaptionskip}{3pt}
	\caption{BER versus the number of iterations (layers) under an \Times{8}{8} MIMO-OFDM system, the WINNER II channel with different condition numbers for 16QAM, \SNR{=}{30}. The number of CG iterations in CG-OAMP/CG-OAMP-NET is $I_{\text{CG}}=50$.}
	\label{fig:cp_free_convergence}
\end{figure}
}

{We also generate channel realizations using the WINNER II model and compute the condition numbers of the generated matrices to investigate the impact of channel conditioning on the competing methods. The channel condition number $\kappa$ is defined as the ratio of the largest to smallest singular value of the channel matrix, which is a metric for evaluating the quality of the MIMO channels \cite{liu_memory_2021}. The evaluated system is an \Times{8}{8} MIMO-OFDM with ${N_c=64}$ subcarriers, which matches the configurations in \tabref{tab:complexity}.
 \figref{fig:cp_free_convergence} shows that all competitive methods experience degradation when $\kappa$ increases. The convergence performance of MAMP is also inferior to OAMP in this realistic channel. By contrast, CG-OAMP-NET still converges in five layers and has a performance advantage over the AMP-type methods regardless of the changing $\kappa$.}

\subsection{Detection Performance in SCP Systems}
We present the detection performance of CG-OAMP-NET in SCP systems
with IID Rayleigh and correlated MIMO channels in this subsection.
The robustness of the network towards SNR and channel correlation mismatches is also demonstrated.

\subsubsection{Rayleigh MIMO Channel Performance}
We first investigate the performance of the CG-OAMP-NET detector under IID Rayleigh MIMO channels.
\figref{fig:rayleigh_QPSK} shows the performance comparison under QPSK.
We consider the following three kinds of system size: ${\text{8}\times\text{8}}$,
${\text{32}\times\text{32}}$, and ${\text{128}\times\text{128}}$ MIMO systems.
\figref{fig:rayleigh_QPSK.sub.1} reveals that MAMP suffers a considerable performance gap compared with the OAMP
in a small- or medium-sized (with the number of antennas less than 100) MIMO system. 
By contrast, the OAMP-NET significantly outperforms the prototypical OAMP algorithm, 
especially in the high SNR regime.
CG-OAMP-NET further exhibits superiority over OAMP-NET in that the complexity
has been markedly reduced without compromising the performance. 
{Moreover, 
the gap between CG-OAMP-NET and the ML solution decreases with the increase in the number of antennas. Specifically, the gaps are approximately 5 and 1 dB if we target at $\text{BER}=10^{-3}$ for the \Times{8}{8} and \Times{32}{32} MIMO systems, respectively.}
\figref{fig:rayleigh_QPSK.sub.2} shows the consistent performance of all the detectors in a relatively large system because OAMP and MAMP are theoretically 
{\BO} for IID Gaussian channels under the large system limit. {We do not provide the ML baseline in \figref{fig:rayleigh_QPSK.sub.2} due to the prohibitive complexity.}

\CheckRmv{
\begin{figure}[tbp]
	\centering
	\subfigure[${\text{8}\times\text{8}}$ MIMO and ${\text{32}\times\text{32}}$ MIMO]{
		\label{fig:rayleigh_QPSK.sub.1}
		\includegraphics[width=3.0in]{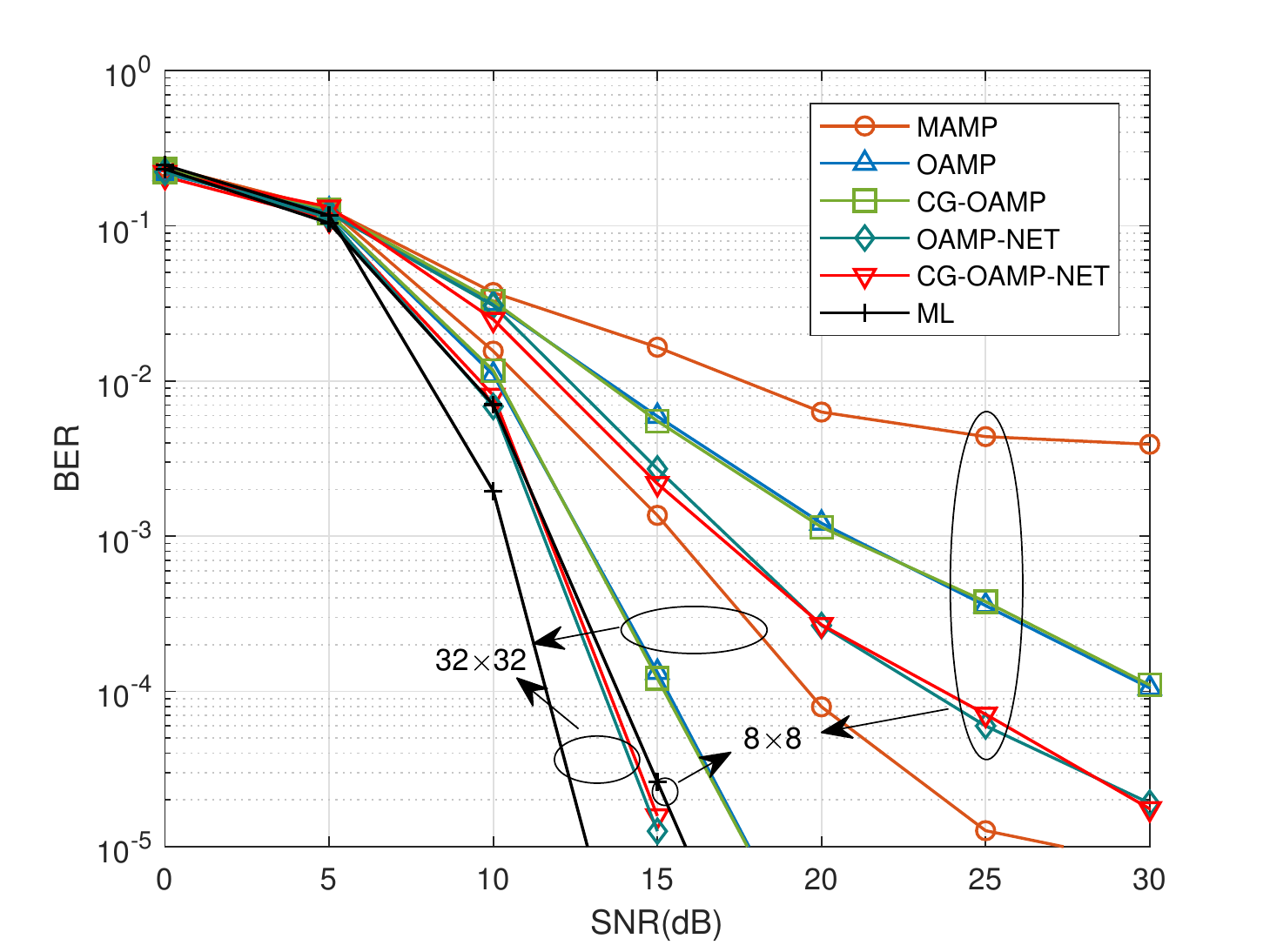}
	}
	\subfigure[${\text{128}\times\text{128}}$ MIMO]{
		\label{fig:rayleigh_QPSK.sub.2}
		\includegraphics[width=3.0in]{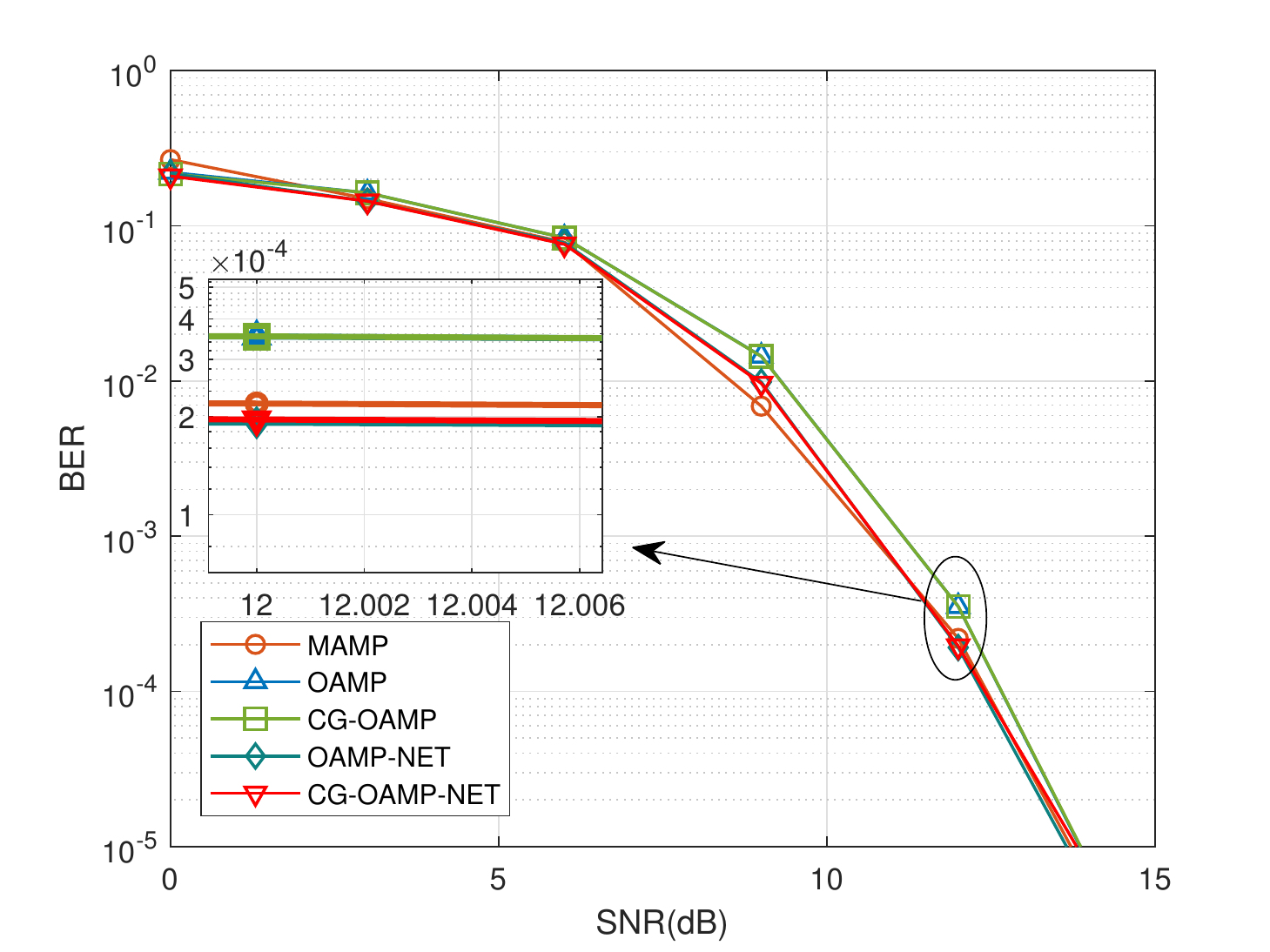}
	}
	\caption{The BER performance comparison under Rayleigh MIMO channels 
	with different antenna configurations for QPSK.}
	\label{fig:rayleigh_QPSK}
\end{figure}
}

\figref{fig:rayleigh_16QAM} also demonstrates the BER performance comparison under high-order modulation, that is, 16QAM. 
The figure reveals that the gain of OAMP-NET and CG-OAMP-NET is still significant for high-order modulation, 
especially for the ${\text{32}\times\text{32}}$ MIMO system. 
Furthermore, we apply the trained parameters with \SNR{=}{30} to other SNRs, 
which is marked by ``Mismatch'' in \figref{fig:rayleigh_16QAM.sub.2}, to investigate the robustness of the CG-OAMP-NET.
Compared with the network trained and tested under the matched SNR value,
the ``Mismatch'' one only has a slight gap in BER, 
which suggests that the CG-OAMP-NET has strong robustness against SNR mismatch.
\CheckRmv{
\begin{figure}[tbp]
	\centering
	\subfigure[${\text{8}\times\text{8}}$ MIMO]{
		\label{fig:rayleigh_16QAM.sub.1}
		\includegraphics[width=3.0in]{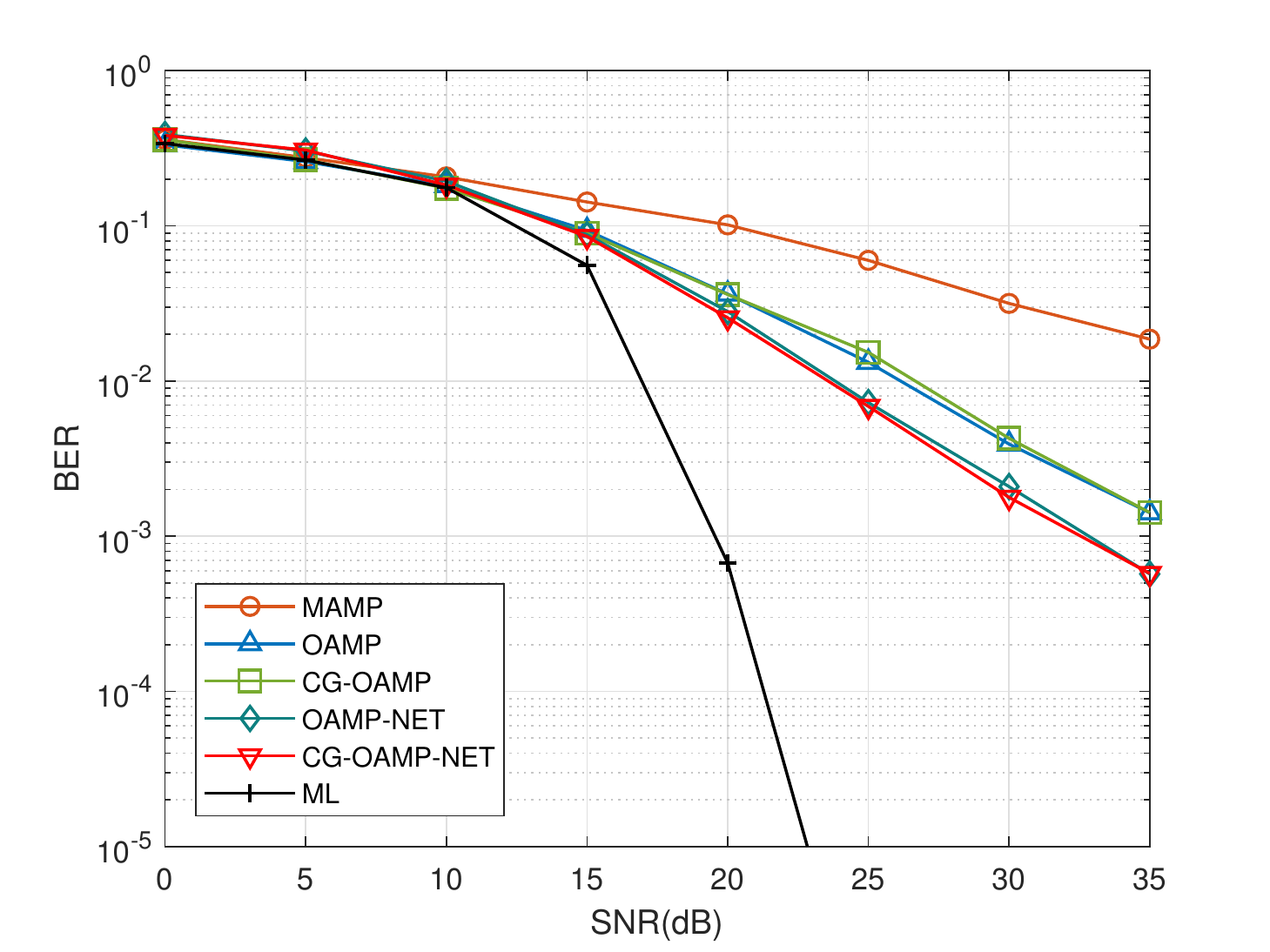}
	}
	\subfigure[${\text{32}\times\text{32}}$ MIMO]{
		\label{fig:rayleigh_16QAM.sub.2}
		\includegraphics[width=3.0in]{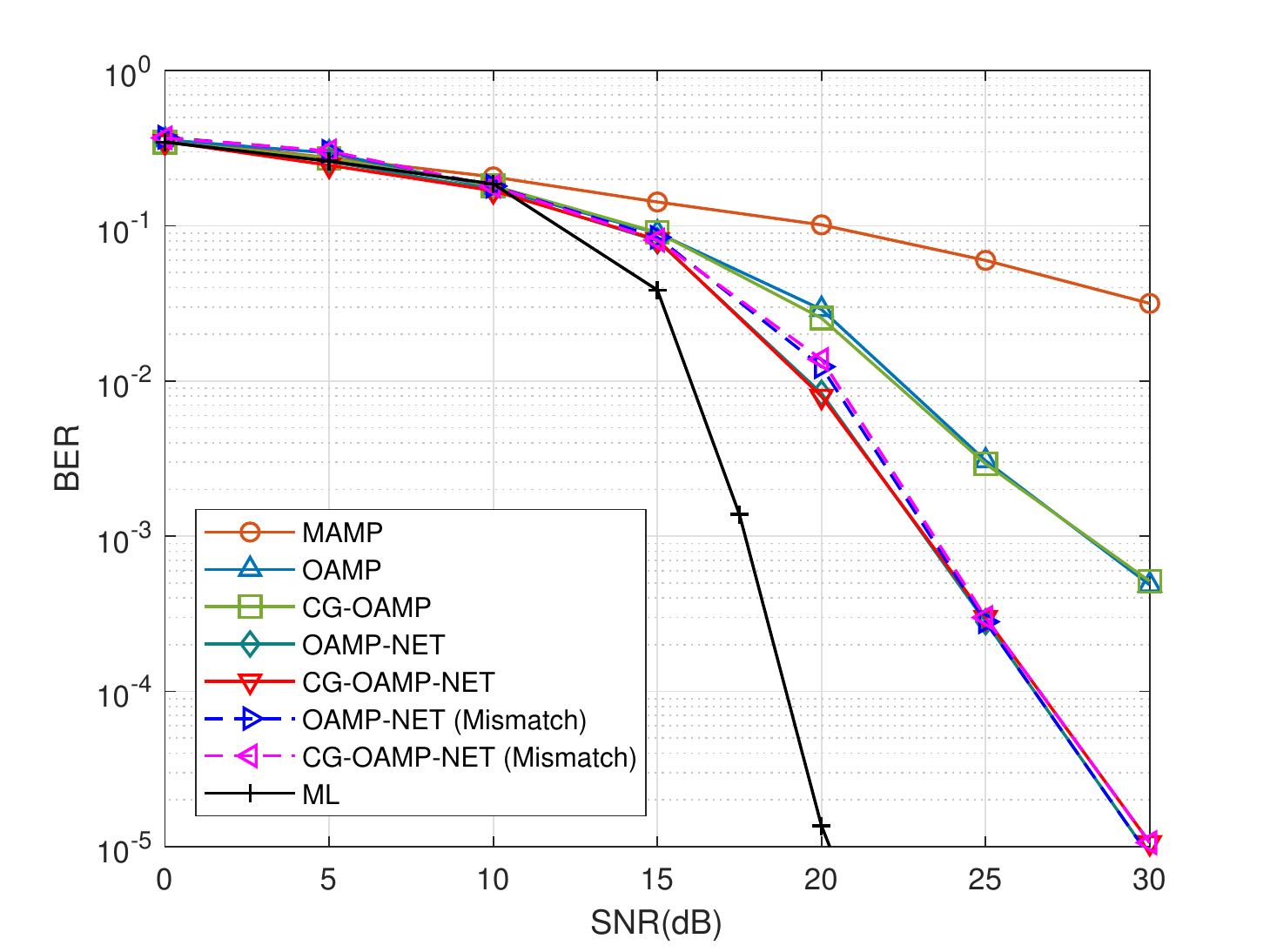}
	}
	\caption{The BER performance comparison under Rayleigh MIMO channels 
	with different antenna configurations for 16QAM.}
	\label{fig:rayleigh_16QAM}
\end{figure}
}
\subsubsection{Correlated MIMO Channel Performance}
We consider the detection performance in a spatially correlated MIMO channel, 
which is a remarkably challenging scenario, in this subsection. 
We use the Kronecker model to describe the correlated channel, 
\CheckRmv{
\begin{equation}
    {\mathbf{G}} = {\mathbf{R}}_r^{1/2}{\mathbf{UR}}_t^{1/2},
    \label{eq:kronecker}
\end{equation}
}
where $\mathbf{U} \in \mathbb{C}^{N_r \times N_t}$ is IID Rayleigh MIMO channel matrix. 
$\mathbf{R}_r \in \mathbb{C}^{N_r \times N_r}$ and $\mathbf{R}_t \in \mathbb{C}^{N_t \times N_t}$
denote the correlation between {\Rx} and {\Tx} antennas, respectively. 
$\mathbf{R}_r$ and $\mathbf{R}_t$ are generated by the exponential correlation model \cite{loyka_channel_2001}
whose element $r_{i,j}$ is given by the following:
\CheckRmv{
\begin{equation}
    r_{i,j}=\left\{
    \begin{array}{ll}
        \rho^{j-i},& i\leq j\\
        r_{j,i}^{\ast},&i>j
    \end{array}\right.,    
\end{equation}
}
where $\rho$ is the correlation coefficient between neighboring antennas. We train the networks with $\rho=0.5$ and
\SNR{=}{20} while testing them with $\rho = 0.5$ or $0.9$ and different SNRs. 
The type of modulation is QPSK, and an \Times{8}{8} MIMO system is considered.
We set the number of layers in OAMP-NET, CG-OAMP-NET, and their prototypical algorithms
as 10 in this situation. 
\figref{fig:correlated} shows that all competitive methods experience degradation to a certain extent
compared with the performance in IID Rayleigh channels. 
However, the proposed CG-OAMP-NET still has performance gain over its prototypical algorithm and MAMP.
This phenomenon can be attributed to the unitarily-invariant channel matrix requirement
of OAMP and MAMP, which cannot be satisfied in a correlated channel.
By contrast, the trainable parameters in CG-OAMP-NET can compensate for the loss caused by channel correlation.
Finally, CG-OAMP-NET still has an advantage over traditional iterative methods when the BER curves with $\rho = 0.9$ 
are targeted, which demonstrates its robustness to correlation coefficient mismatch.
 
\CheckRmv{
\begin{figure}[tbp]
\setlength{\abovecaptionskip}{-0.3cm}
\setlength{\belowcaptionskip}{-0.0cm}
	\centerline{\includegraphics[width=3.0in]{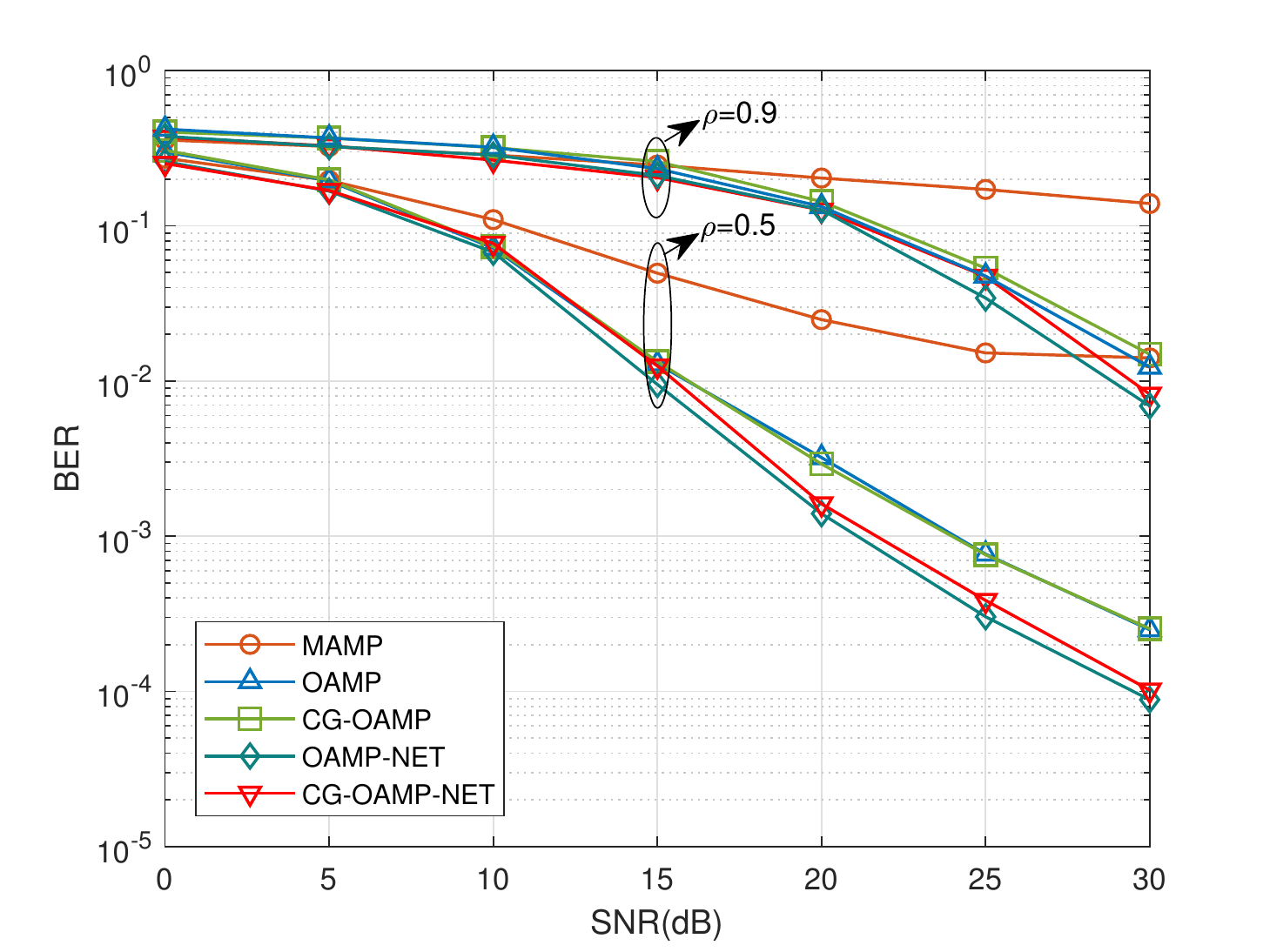}}
	\caption{The BER performance comparison under correlated MIMO channels 
	with $\rho=0.5$ and $0.9$ for QPSK modulation.}
	\label{fig:correlated}
\end{figure}
}

\subsection{Comparison of SCP and CP-Free Schemes {with WINNER II channels}} \label{sec:MIMO_OFDM}
We first present the BER performance, then compare the spectral efficiency of the SCP and CP-free MIMO-OFDM systems, {and finally evaluate the robustness of the algorithms under an imperfect CSI} in this section.
The channel model is WINNER II, and {a \Times{20}{16} MIMO configuration with 16QAM is considered for this comparison. We equip the receiver with more antennas to improve the detection performance under realistic channels.}
We use the receiver architecture developed in Section \ref{sec:CP_free} for the CP-free system {(i.e., all the competitive detectors use the same ISI-reduction scheme)}.
The networks in this comparison are trained with \SNR{=}{20}.

\subsubsection{BER Performance}
The BER performance of the CG-OAMP-NET detector and the competing schemes
is shown in \figref{fig:MIMO_OFDM_BER}. The curves with the prefix
``CP-'' denote the BER performance in the traditional SCP MIMO-OFDM system; 
otherwise, they denote those in the CP-free system. 
The figure shows that the model-driven CG-OAMP-NET still outperforms
conventional AMP-type algorithms and has no performance loss compared with OAMP-NET.
Moreover, the proposed method for CP-free transmission has effectively mitigated the ICI and ISI
caused by CP removal, and the performance approaches that of the receiver with adequate CP.

\subsubsection{Spectral Efficiency}
We use the definition of spectral efficiency as
\CheckRmv{
\begin{equation}
	\text{Spectral efficiency}
	=\frac{\text{Number of correctly received bits}}
	{\text{Total time}\times\text{Total bandwidth}}
	= \frac{\xi R_b(1-\textrm{BER})}{B} (\text{bps/Hz}),
	\label{eq:SE} 
\end{equation}
}
where $R_b$ is the total bit rate, $B$ is the system bandwidth, and the
channel utilization ratio $\xi$ of SCP and CP-free systems 
can be defined as \cite{liu_symbol_2017}
\CheckRmv{
\begin{equation}
    \xi=\left\{
    \begin{array}{cl}
        1, & \text{CP-free}\\
        \displaystyle\frac{N_c}{N_c+N_g},&\text{SCP}
    \end{array} \right..   
\end{equation}
} 

We derive the results in \figref{fig:MIMO_OFDM_SE} by combining the BER performance in 
\figref{fig:MIMO_OFDM_BER} and the definition of spectral efficiency in \eqref{eq:SE}.
\figref{fig:MIMO_OFDM_SE} demonstrates that the CP-free scheme eliminates the extra spectrum overhead consumed
by CP transmission, substantially improving the spectral efficiency compared to the
traditional scheme with CP.

\CheckRmv{
\begin{figure}[t]
	\centering
	\subfigure[BER]{
		\label{fig:MIMO_OFDM_BER}
		\includegraphics[width=3.0in]{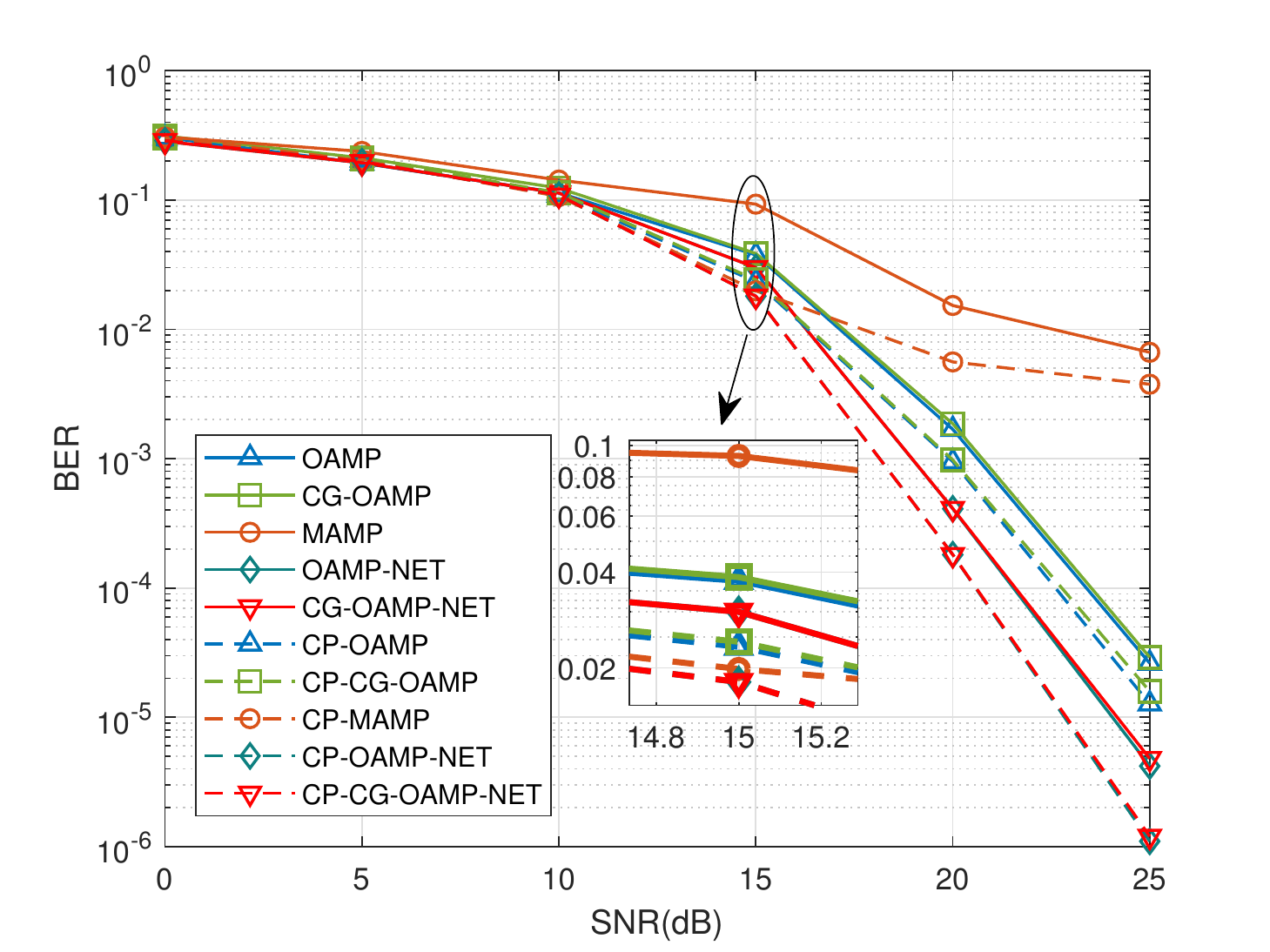}
	}
	\subfigure[Spectral efficiency]{
		\label{fig:MIMO_OFDM_SE}
		\includegraphics[width=3.0in]{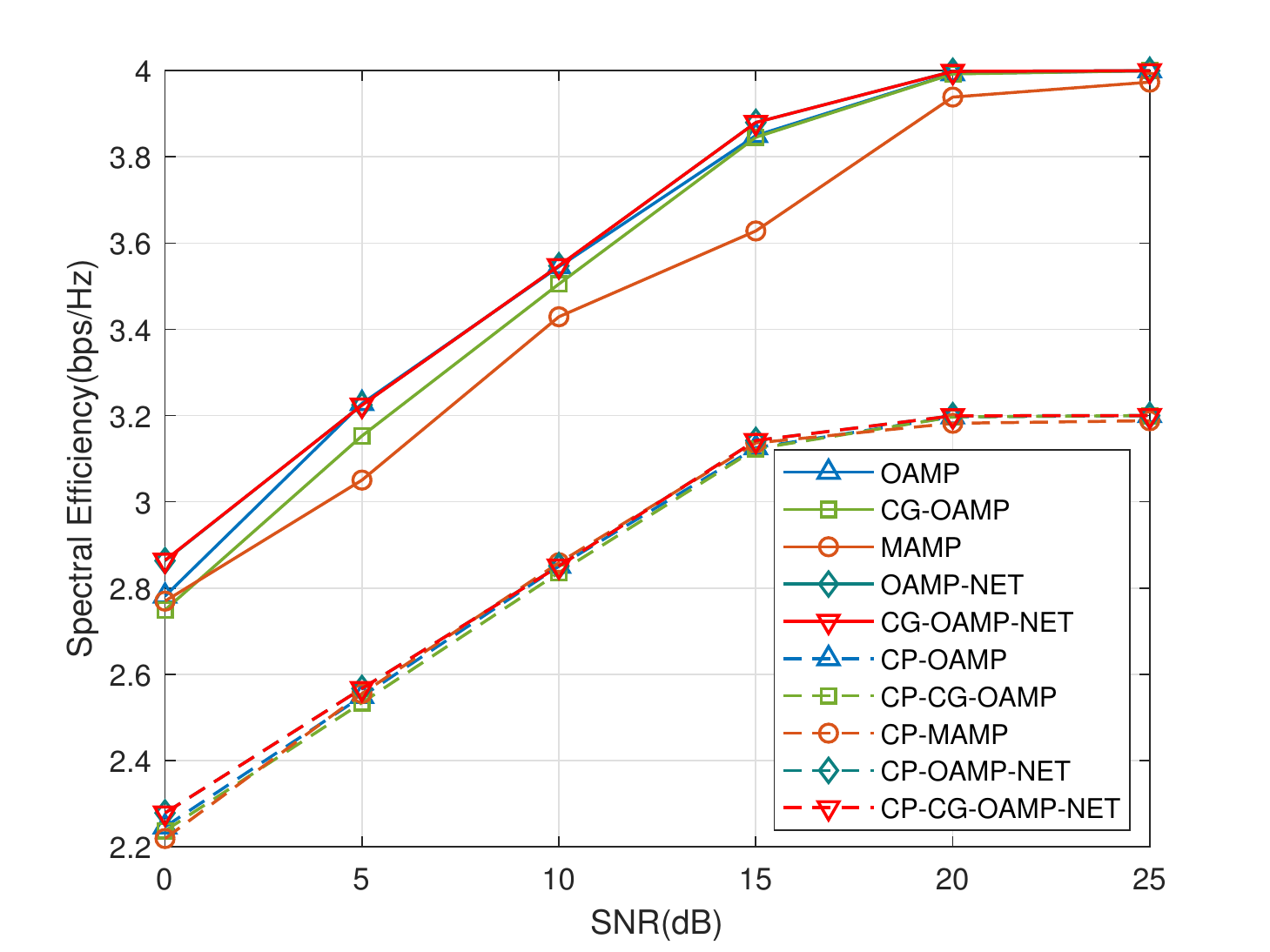}
	}
	\caption{BER and spectral efficiency of MIMO-OFDM systems
	with and without CP under 16QAM modulation.}
	\label{fig:MIMO_OFDM}
\end{figure}
}
{\subsubsection{Imperfect CSI} 

In the above, all detectors are investigated with an accurate CSI. 
We test the CG-OAMP-NET with an imperfect CSI under the aforementioned configurations to verify the robustness of the algorithm against channel estimation errors in this subsection. 
Specifically, we assume that one pilot and six data OFDM blocks constitute one frame. The channel is estimated either by the LMMSE method \cite{li_robust_1998} or by the channel estimation network (CE-NET) proposed in \cite{gao_comnet_2018} to overcome the nonlinear effect in the CP-free system. 
{The CE-NET uses the least-square method to provide an initial estimation and refine the estimated channels by a two-layer NN.}
 \figref{fig:CSIerror} shows the BER performance under perfect CSI and estimated CSIs using LMMSE and CE-NET, marked by the prefixes ``CSI-'', ``LMMSE-'', and ``CE-NET-'', respectively. We do not show the BER curve of MAMP because the performance becomes even worse with an imperfect CSI.
We observe a performance gap between perfect and estimated CSIs, demonstrating that the CSI accuracy affects the performance of all the tested algorithms. 
For example, the channel estimation accuracy of LMMSE significantly drops under the CP-free case, resulting in the pitfalls of all detectors. However, 
the CG-OAMP-NET always has a performance advantage over the OAMP under different channel estimation errors.  Meanwhile, the CG-OAMP-NET with an estimated CSI using CE-NET performs as well as OAMP with a perfect CSI, demonstrating the robustness of the proposed algorithm against channel estimation errors.
}

\CheckRmv{\begin{figure}[tbp]
	\centering
	\subfigure[SCP]{
	\centering
	\label{fig:cp_csi}
	\includegraphics[width=3in]{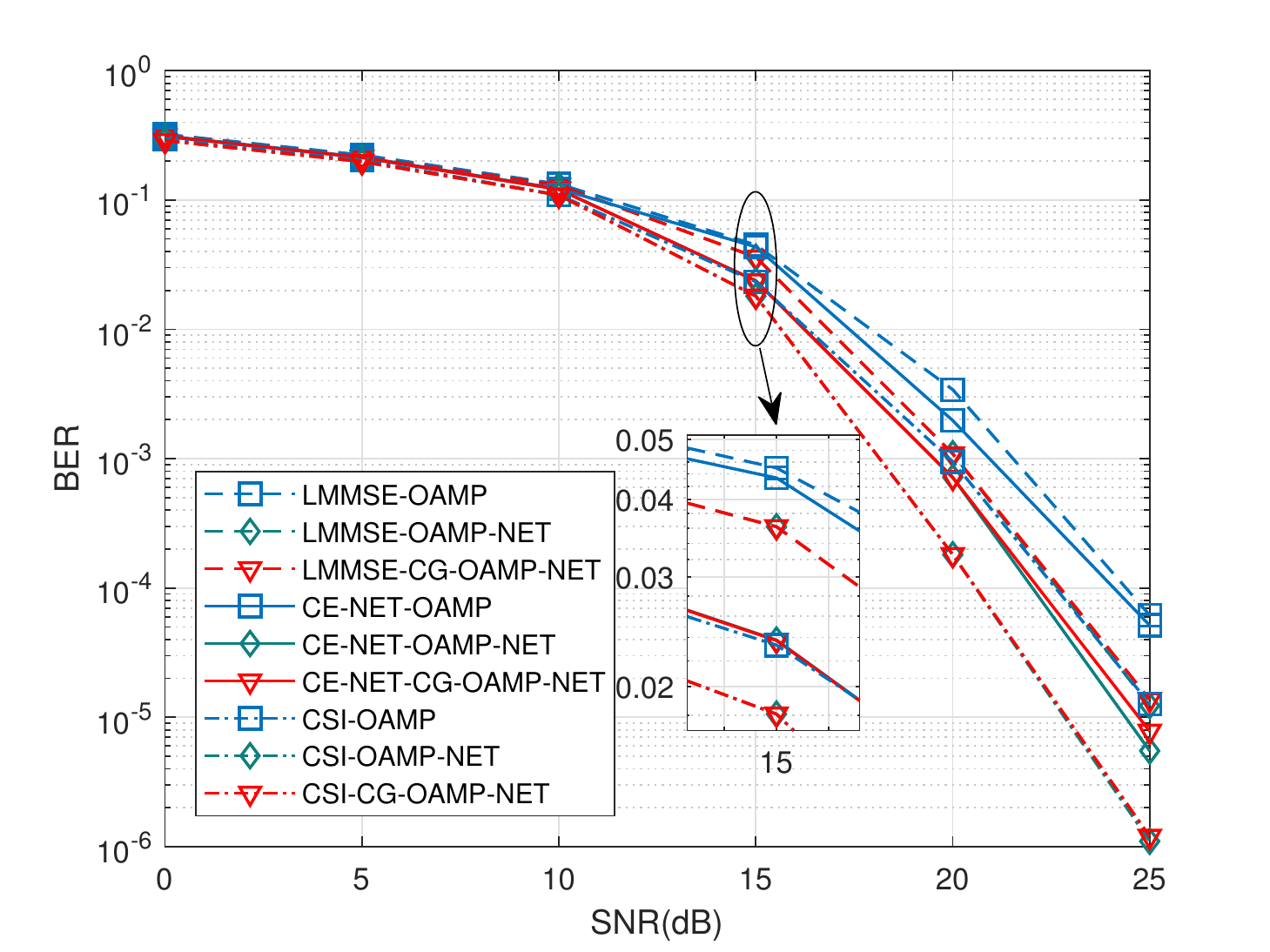}
	}
	\subfigure[CP-free]{
	\label{fig:cp_free_csi}
	\centering
	\includegraphics[width=3in]{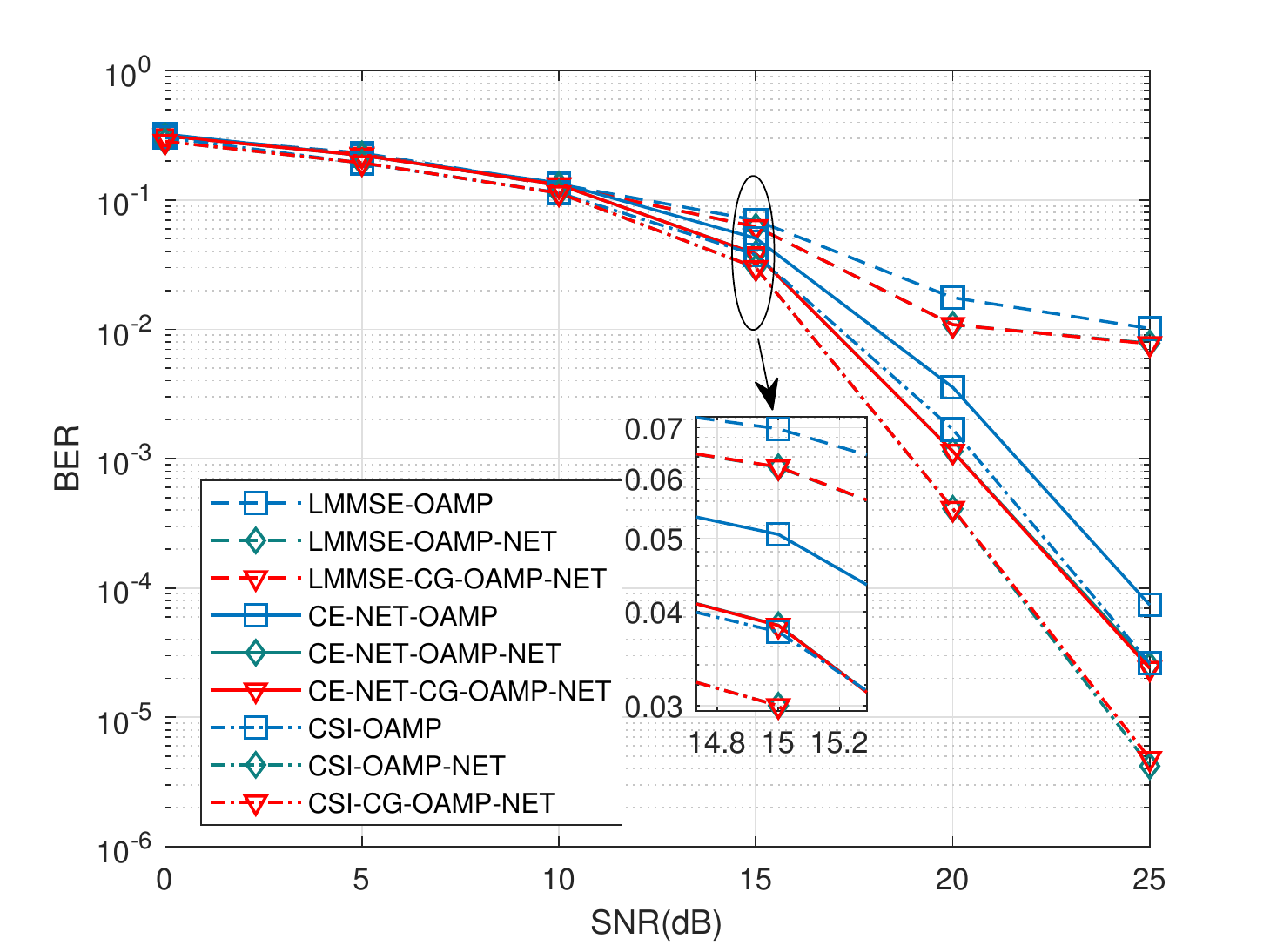}
	}
	\caption{BER performance comparison between perfect and estimated CSIs.}
	\label{fig:CSIerror}
\end{figure}}
		
\section{OTA Test and Discussion} \label{sec:OTA}
An OTA platform is built in this section to validate the effectiveness and 
robustness of the proposed algorithm in realistic propagation scenarios.

\vspace{-0.5cm}
\subsection{System Setup}
\subsubsection{Prototyping System Architecture}
\figref{fig:arch} shows the architecture of the prototyping system, which is a
${\text{20}\times\text{16}}$  MIMO-OFDM platform operating in the 3.5 GHz band 
(3400--3600 MHz) \cite{tsai_experimental_2018}.
Two distributed eight-port patch antennas are used as the transmitter,
which simultaneously sends 16 MIMO symbol streams. The receiver comprises two smartphones
each embedded with 10 {\Rx} antennas, which are equally integrated along the two
long sides of the equipment.  {We equip more antennas on the compact smartphones to enhance the equivalent receiving SNR considering the high correlation between multiple antennas on the limited space.}
The transmitter and receiver configurations are comprehensively described as follows:
\begin{itemize}
	\item Transmitter configurations: At the transmitter, the Rohde and Schwarz (R\&S)
	signal generator (SGT100A) is used to generate modulated radio frequency (RF) signal. 
	The baseband transmission uses a coded MIMO-OFDM mechanism, 
	which is different from the numerical simulations. The information bits are 
	encoded by a convolutional encoder with generator polynomial $[{133}_o\,{171}_o]$ 
	and code rate ${R=1/2}$. QPSK and 64QAM modulations are conducted.
	Afterward, the signal waveform is shaped and then up-converted to the 5G frequency
	band of 3.5 GHz. Finally, the {\Tx} antennas emit the RF signal into the wireless channel.
	\item Receiver configurations: At the receiver, {each antenna connects to one RF chain}. The R\&S oscilloscope (RTO2044) 
	works as a down converter, which first converts the RF analog signal into the digital 
	domain and then moves it to the baseband. The baseband signal is 
	then sent to a software receiver for processing via a local area network. 
	The baseband signal processing of the software receiver contains signal synchronization, 
	channel estimation, MIMO detection, and channel decoding to recover the raw bits, where different detectors are implemented for comparison, and the Viterbi algorithm is chosen for decoding. 
	In particular, the model-driven DL-based detectors are first trained offline and
	then deployed on a computer with an Intel CPU (1.60 GHz) and 16 GB memory
	for online detection.
\end{itemize}

\CheckRmv{
\begin{figure}[t]
	\setlength{\abovecaptionskip}{0.0cm}
	\centerline{\includegraphics[width=4.5in]{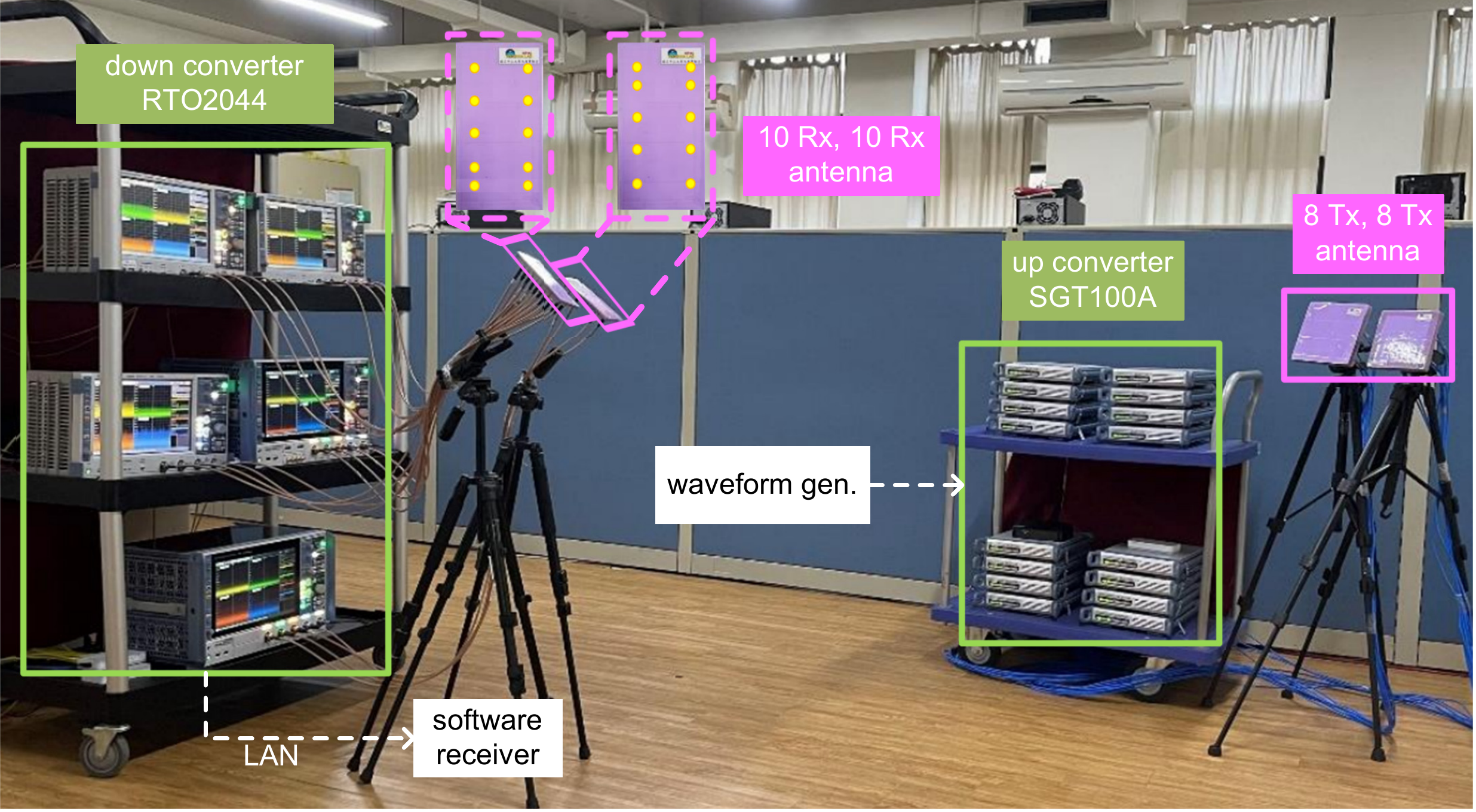}}
	\caption{The ${\text{20}\times\text{16}}$ MIMO-OFDM prototyping platform. 
	}
	\label{fig:arch}
\end{figure}
}

\subsubsection{Frame Structure}

\CheckRmv{
\begin{figure}[tbp]
\setlength{\abovecaptionskip}{-0.2cm}
\setlength{\belowcaptionskip}{-0.0cm}
	\centerline{\includegraphics[width=6.0in]{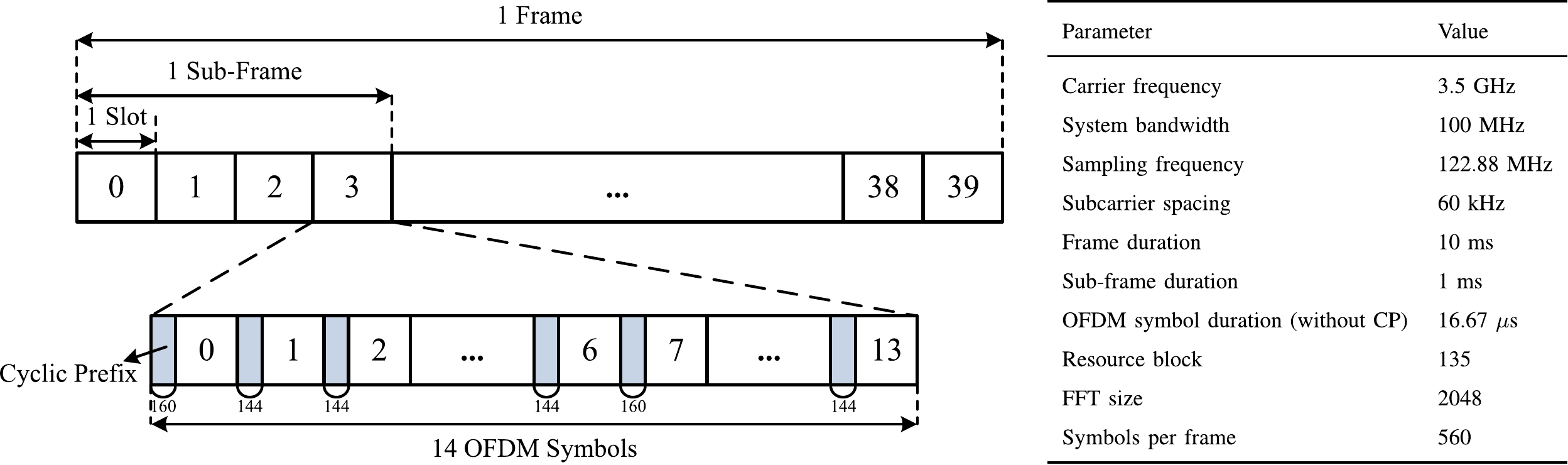}}
	\caption{5G NR Frame Structure.}
	\label{fig:frame}
\end{figure}
}
The prototyping system is based on the 5G New Radio (NR) frame structure (Rel.15).
The system bandwidth is set to 100 MHz, and the complete frame duration is 10 ms.
\figref{fig:frame} shows 10 sub-frames of equal length within one frame,
and each sub-frame comprises four time slots. The number of OFDM symbols per slot is 14. 
In every 7 symbols, the first symbol has a CP of 160 sampling
points, while the CP length for other symbols is 144 sampling points.
{Moreover, 12 subcarriers and 14-time units (a slot) constitute a resource block in the transmission. 
In each resource block, two pilots are inserted into the frequency and time domains, respectively, to facilitate channel estimation. 
First, the channel coefficients in the frequency domain are derived by utilizing the LMMSE method \cite{li_robust_1998}. Then, we use linear interpolation to obtain the channel estimation in the time domain. }
The transmission parameters are also organized in \figref{fig:frame}. Notably,
although 2048 points FFT are used for transmission, we only extract the first 64 effective 
subcarriers for symbol detection and performance evaluation. 

\subsection{Scenario Description and Algorithm Implementation}
\CheckRmv{
\begin{figure}[t]
\setlength{\abovecaptionskip}{-0.3cm}
\setlength{\belowcaptionskip}{-0.0cm}
	\centerline{\includegraphics[width=4in]{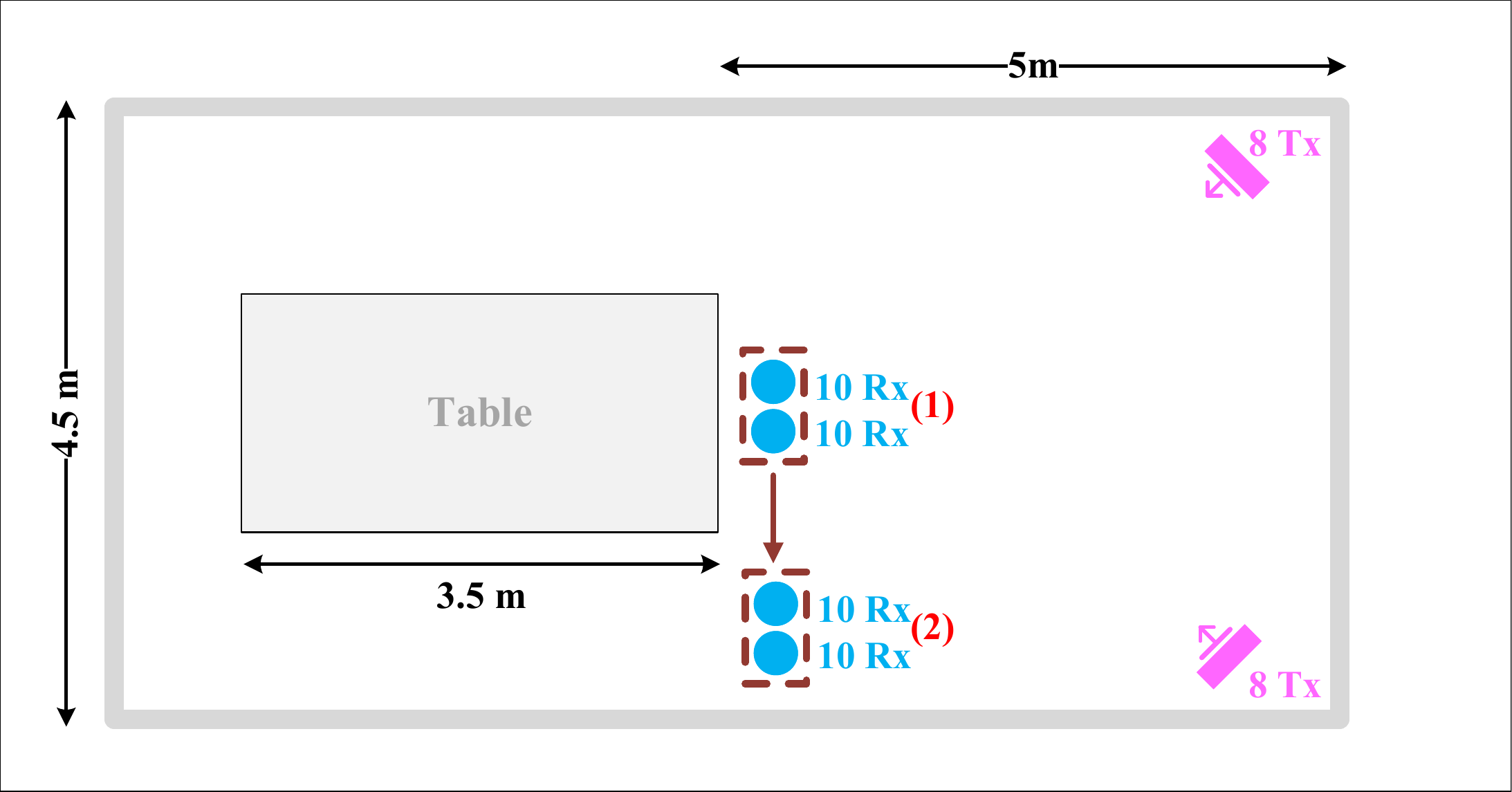}}
	\caption{Diagram of the office where the measurements
	are performed.}
	\label{fig:scenario}
\end{figure}
}

\CheckRmv{
\begin{table*}[t]
	\caption{Measurement scenarios}
	\begin{center}
	\begin{tabular}{l|l}
		\toprule
		Cases& Description \\
		\midrule
		i. RX (1) & Place the receiver at position (1) and collect data as the benchmark. \\
		ii. Walking & Place the receiver in the same position as the benchmark, with a few people walking.\\
		iii. 30min & Place the receiver in the same position as the benchmark and take the measurement after 30 min.\\
		iv. Angle 1 & Change the angle of the receiver antenna at the same location as the benchmark. \\ 
		v. Angle 2 &  Change to another receiving angle different from iv. at the same location as the benchmark.\\
		vi. RX (2) & Move the receiver to position (2) and collect data in a different scenario. \\
		\bottomrule
	\end{tabular}
	\label{tab:OTA case}
	\end{center}
\end{table*}
}
Measurements with the aforementioned system setup are conducted in an indoor office,
where the receiver is arranged in different locations to collect data at different times.
\figref{fig:scenario} presents the floor plan of the office where the measurements are
performed. The measurements are divided into several cases, and the main cases 
are listed in \tabref{tab:OTA case}. In this table, one case is chosen as the benchmark, and 
the others are based on the corresponding modification to the benchmark. 

In the practical test, the OAMP and CG-OAMP as well as the corresponding networks
run for 10 iterations before stopping. The number of CG iterations is 
${I_{\text{CG}}=50}$ and ${I_{\text{CG}}=200}$ for QPSK and 64QAM, respectively. 
Meanwhile, the number of iterations in MAMP is 50. We train the detection NNs
under the benchmark (case i), and the NNs are tested in other scenarios 
(cases ii to vi) after obtaining the trained parameters to verify their robustness to mismatches. 
Similar to the simulation, we study the performance comparison between 
SCP and CP-free schemes.

\subsection{Experimental Results}
Tables~\ref{tab:OTA_BER_CP} and \ref{tab:OTA_BER_CP_free} provide the experimental
results in the prototyping system with and without CP, respectively.
The following observations are obtained from the results.

\CheckRmv{
\begin{table*}[tbp]
	\caption{BER performance for MIMO-OFDM system in the measurement scenarios}
	\begin{center}
	\begin{tabular}{llc|c|c|c|c|c}
		\toprule
		Modulation & Algorithm & i. RX (1) & ii. Walking & iii. 30 min & iv. Angle 1 & v. Angle 2 & vi. RX (2)  \\
		\midrule
		\multirow{5}{*}{QPSK} & OAMP & 1.74e-4 & 1.74e-3 & 6.45e-5 & 2.47e-4 & 5.69e-4 & 4.27e-5\\
		& CG-OAMP & 1.75e-4 & 1.74e-3 & 6.54e-5 & 2.47e-4 & 5.51e-4 & 4.32e-5\\
		& MAMP & 2.61e-2 & 3.52e-2 & 1.73e-2 & 1.55e-2 & 2.61e-2 & 6.71e-3 \\
		& OAMP-NET & \textbf{5.60e-6} & \textbf{1.99e-4} & \textbf{3.30e-6} & \textbf{1.90e-5} & \textbf{3.16e-5} & \textbf{9.00e-7}\\
		& CG-OAMP-NET & \textbf{5.71e-6} & \textbf{1.84e-4} & \textbf{3.30e-6} & \textbf{1.91e-5} & \textbf{3.20e-5} & \textbf{1.40e-6}\\
		\midrule
		\multirow{5}{*}{64QAM} & OAMP & 2.47e-3 & 2.46e-3 & 5.32e-4 & 1.99e-4 & 1.05e-3 & 3.28e-4\\
		& CG-OAMP & 2.48e-3 & 2.46e-3 & 5.86e-4 & 1.90e-4 & 1.17e-3 & 3.50e-4\\
		& MAMP & 3.82e-1 & 3.36e-1 & 3.65e-1 & 3.77e-1 & 3.89e-1 & 3.62e-1 \\
		& OAMP-NET & \textbf{1.07e-3} & \textbf{1.35e-3} & \textbf{1.50e-4} & \textbf{1.44e-4} & \textbf{6.12e-4} & \textbf{1.23e-4}\\
		& CG-OAMP-NET & \textbf{1.24e-3} & \textbf{1.52e-3} & \textbf{1.76e-4} & \textbf{1.38e-4} & \textbf{7.59e-4} & \textbf{1.19e-4}\\
		\bottomrule
	\end{tabular}
	\label{tab:OTA_BER_CP}
	\end{center}
\end{table*}
}

\CheckRmv{
\begin{table*}[tbp]
	\caption{BER performance for CP-free MIMO-OFDM system in the measurement scenarios}
	\begin{center}
	\begin{tabular}{llc|c|c|c|c|c}
		\toprule
		Modulation & Algorithm & i. RX (1) & ii. Walking & iii. 30 min & iv. Angle 1 & v. Angle 2 & vi. RX (2)  \\
		\midrule
		\multirow{5}{*}{QPSK} & OAMP & 8.36e-3 & 6.35e-3 & 8.10e-3 & 1.02e-2 & 8.72e-3 & 1.56e-3\\
		& CG-OAMP & 8.36e-3 & 6.60e-3 & 7.83e-3 & 1.02e-2 & 8.76e-3 & 1.56e-3\\
		& MAMP & 4.89e-2 & 5.82e-2 & 3.98e-2 & 3.93e-2 & 5.74e-2 & 1.86e-2 \\
		& OAMP-NET & \textbf{2.85e-3} & \textbf{3.40e-3} & \textbf{3.85e-3} & \textbf{5.70e-3} & \textbf{3.47e-3} & \textbf{6.17e-4}\\
		& CG-OAMP-NET & \textbf{2.91e-3} & \textbf{3.62e-3} & \textbf{3.27e-3} & \textbf{5.87e-3} & \textbf{3.07e-3} & \textbf{6.02e-4}\\
		\midrule
		\multirow{5}{*}{64QAM} & OAMP & 1.43e-2 & 8.12e-3 & 9.79e-3 & 6.39e-3 & 1.20e-2 & 7.94e-3\\
		& CG-OAMP & 1.37e-2 & 9.33e-3 & 8.68e-3 & 6.42e-3 & 1.24e-2 & 8.22e-3\\
		& MAMP & 4.27e-1 & 4.23e-1 & 3.99e-1 & 3.91e-1 & 4.25e-1 & 3.96e-1 \\
		& OAMP-NET & \textbf{5.08e-3} & \textbf{4.15e-3} & \textbf{3.67e-3} & \textbf{3.62e-3} & \textbf{5.46e-3} & \textbf{2.80e-3}\\
		& CG-OAMP-NET & \textbf{5.45e-3} & \textbf{4.18e-3} & \textbf{3.78e-3} & \textbf{3.78e-3} & \textbf{5.50e-3} & \textbf{2.82e-3}\\
		\bottomrule
	\end{tabular}
	\label{tab:OTA_BER_CP_free}
	\end{center}
\end{table*}
}

\begin{itemize}
	\item MAMP performs poorly in the OTA test and cannot nearly work for high-order modulations.
	This phenomenon is consistent with the simulation results, which can still be
	explained by the spatial correlation in real-life channels, breaking the 
	unitarily-invariant prerequisite.
	\item The BER performance is quite different when the receiver is located at different places
	 or with different orientations given a detection algorithm. Taking the QPSK modulation as 
	an example, case vi has considerable gain over the benchmark because the location is different, and the performance of cases iv and v is not as good as that of the benchmark
	due to the modified receiving angle. 
	Moreover, case iii has a similar performance to the benchmark, which means that the 
	indoor environment is nearly time-invariant. 
	\item {Among all tested detectors, the model-driven DL-based OAMP-NET and CG-OAMP-NET have 
	the best performance because they can learn the appropriate parameters for realistic 
	propagation scenarios.} Considering the BER performance under cases ii to vi,
	the proposed algorithm still has a significant improvement over the conventional
	AMP-type methods despite the difference of the environment from that in the
	training phase. This finding demonstrates that CG-OAMP-NET exhibits substantial robustness
	against environmental changes and can avoid frequent re-training.
	Thus, the proposed CG-OAMP-NET can be regarded as a high-performance and efficient solution
	for detection in practical MIMO-OFDM systems considering the complexity analysis in Section \ref{sec:complexity}.
	\item  Compared with the baseline with adequate CP, all algorithms suffer a performance loss 
	in real-life CP-free systems despite ICI and ISI suppression to a large extent by the proposed scheme due to 
	the long delay spread and the severe interference.
	Nevertheless, the model-driven NNs still perform better than other schemes in CP-free scenarios.
\end{itemize}

\vspace{-0.5cm}
\section{Conclusions} \label{sec:conclusion} 
\vspace{-0.2cm}
We proposed a model-driven DL-based detector named CG-OAMP-NET for MIMO-OFDM systems. 
The proposed approach stemmed from the OAMP detector, and the detector was revised by
using CG to replace matrix inversion. 
We then unfolded the revised detector into a network and added some trainable parameters. {These parameters could be tuned through the DL techniques to enhance detection performance.}
Complexity analysis and simulation results showed that the proposed CG-OAMP-NET 
effectively reduced the cost of OAMP while achieving remarkable detection performance.
Thus, a desirable tradeoff between complexity and performance could be attained.
Furthermore, we applied the CG-OAMP-NET to the design of a CP-free MIMO-OFDM receiver.
Numerical and OTA experiments confirmed that the proposed receiver is an efficient
scheme for MIMO-OFDM because it can minimize spectrum loss.
The experiments also verified the robustness of the proposed scheme in realistic scenarios. {The WINNER II and OTA channel datasets used in this work are available at \url{https://github.com/STARainZ/CG-OAMP-NET}.}

\vspace{-0.2cm}
\appendix[The derivation for $\tau^2$ in \eqref{eq:tau2} and \eqref{eq:tau2_NET}]  \label{sec:appendix}
For convenience, the subscript index $t$ is omitted in this derivation. We first consider 
the expression \eqref{eq:tau2} for $\tau^2$ in OAMP. The trace $\tr(\mathbf{BB}^T)$ in 
\eqref{eq:tau2_old} can be expressed as 
\CheckRmv{  
\begin{align}
	{\tr}(\mathbf{BB}^{T})=&{\tr}((\mathbf{I}-\mathbf{W}\bar{\mathbf{G}})
	(\mathbf{I}- \bar{\mathbf{G}}^T\mathbf{W}^{T})) \nonumber  \\
	=&2N_t-{\tr}(\mathbf{W}\bar{\mathbf{G}})
	-{\tr}(\bar{\mathbf{G}}^{T}\mathbf{W}^T)+{\tr}(\mathbf{W}\bar{\mathbf{G}}\bar{\mathbf{G}}^{T} \mathbf{W}^{T})\nonumber  \\
	=&-2N_t+{\tr}(\mathbf{W}\bar{\mathbf{G}}\bar{\mathbf{G}}^{T} \mathbf{W}^{T}),
\label{eq:trBBT}
\end{align}
}
where the last equality originates from the de-correlated property of $\mathbf{W}$,
that is, ${\tr}({\bf{W}}\bar{\mathbf{G}})= {\tr}({\bar{\mathbf{G}}^T}{{\bf{W}}^T}) = 
{\tr}(\frac{{2N_t}}{{{\tr}(\hat{\bf{W}}\bar{\mathbf{G}})}} \hat{\bf{W}}\bar{\mathbf{G}}) = 2N_t$.
Substituting \eqref{eq:trBBT} into \eqref{eq:tau2_old}, we derive
\CheckRmv{ 
\begin{equation}
	{\tau ^2} = -{v^2}+\frac{v^{2}}{2N_t} 
	{\tr}(\mathbf{W} \bar{\mathbf{G}}\bar{\mathbf{G}}^{T} \mathbf{W}^{T})
	+\frac{ {\sigma_{\nu}^2}}{4N_t} {\tr}(\mathbf{W} \mathbf{W}^{T}).
\label{eq:tau2_appendix}
\end{equation}
}
Considering the summation of the second and third terms in \eqref{eq:tau2_appendix},
\CheckRmv{ 
\begin{align}
	\frac{v^{2}}{2N_t}{\tr}(\mathbf{W} \bar{\mathbf{G}}\bar{\mathbf{G}}^{T} \mathbf{W}^{T})
	+\frac{ {\sigma_{\nu}^2}}{4N_t} {\tr}(\mathbf{W} \mathbf{W}^{T})
	&\overset{\text{(a)}}{=}\frac{v^2\zeta^2}{2N_t}\Big({\tr}(\hat{\mathbf{W}} \bar{\mathbf{G}}\bar{\mathbf{G}}^{T} \hat{\mathbf{W}}^{T})
	+\frac{\sigma_{\nu}^2}{2v^2} {\tr}(\hat{\mathbf{W}} \hat{\mathbf{W}}^{T}) \Big)\nonumber\\ 
	&\overset{\text{(b)}}{=}\frac{v^2\zeta^2}{2N_t}{\tr}(\hat{\mathbf{W}}^{T} \bar{\mathbf{G}}^{T}
	(\bar{\mathbf{G}}\bar{\mathbf{G}}^{T}+\frac{\sigma_{\nu}^2}{2v^2}\mathbf{I} )^{-1}  
	(\bar{\mathbf{G}}\bar{\mathbf{G}}^{T}+\frac{\sigma_{\nu}^2}{2v^2}\mathbf{I} )) \nonumber\\
	&=\frac{v^2\zeta^2}{2N_t}{\tr}(\hat{\mathbf{W}}^{T} \bar{\mathbf{G}}^{T})
	=\frac{v^2\zeta^2}{2N_t}\frac{2N_t}{\zeta}=v^2\zeta,
	\label{eq:tau2_appendix23}
\end{align}
}
where $\overset{\text{(a)}}{=}$ and $\overset{\text{(b)}}{=}$ originate from the definition of 
$\mathbf{W}$ in \eqref{eq:W} and \eqref{eq:LMMSE}, respectively. Thus, combining
\eqref{eq:tau2_appendix} and \eqref{eq:tau2_appendix23}, \eqref{eq:tau2} is derived.

Deriving the expression of the error variance estimator for CG-OAMP-NET in 
\eqref{eq:tau2_NET} is just a special case of the above derivation for \eqref{eq:tau2},
where a trainable parameter $\theta$ is added into \eqref{eq:tau2_old} according
to \cite[(46)]{he_model-driven_2020} to get
\CheckRmv{
\begin{equation}
	\tau^2 = \frac{{v^2}}{{2}N_t}\tr(\hat{\mathbf{B}}\hat{\mathbf{B}}^T)+
	\frac{\theta^2\sigma_{\nu}^2}{{4}N_t}\tr(\mathbf{WW}^T),
	\label{eq:tau2_NET_old}
\end{equation}
}
where $\hat{\mathbf{B}}=\mathbf{I}-\theta \mathbf{W}\bar{\mathbf{G}}$ is the revised version
of ${\mathbf{B}}$. Then, \eqref{eq:trBBT} and \eqref{eq:tau2_appendix} become
\CheckRmv{  
\begin{align}
	{\tr}(\hat{\mathbf{B}}{\hat{\mathbf{B}}}^T)&=2N_t(1-2\theta)+\theta^2 {\tr}
	(\mathbf{W}\bar{\mathbf{G}}\bar{\mathbf{G}}^{T} \mathbf{W}^{T}),\\
	{\tau ^2} &= {v^2}(1 - 2\theta )+\frac{\theta^{2} v^{2}}{2N_t} 
	{\tr}(\mathbf{W} \bar{\mathbf{G}}\bar{\mathbf{G}}^{T} \mathbf{W}^{T})
	+\frac{\theta^{2} {\sigma_{\nu}^2}}{4N_t} {\tr}(\mathbf{W} \mathbf{W}^{T}). \label{eq:tau2_NET_appendix23}
\end{align}
} 
The summation of the second and third terms in \eqref{eq:tau2_NET_appendix23} yields
${v^2}{\theta ^2}\zeta$. Thus, \eqref{eq:tau2_NET} is derived.



\ifCLASSOPTIONcaptionsoff
  \newpage
\fi



%


\end{document}